\documentclass[fleqn,usenatbib]{mnras}

\usepackage{newtxtext,newtxmath}

\usepackage[T1]{fontenc}
\usepackage{ae,aecompl}

\usepackage{graphicx}	
\usepackage{amsmath}	
\usepackage{amssymb}	

\newcommand{\be}{\begin{equation}}
\newcommand{\ee}{\end{equation}}
\newcommand{\ba}{\begin{eqnarray}}
\newcommand{\ea}{\end{eqnarray}}
\newcommand{\vect}[1]{{\boldsymbol{#1}}}
\newcommand{\cha}[1]{\textbf{#1}}

\title[3D anisotropy of velocity fluctuations]{Three-dimensional local anisotropy of velocity fluctuations in the solar wind}

\author[A. Verdini et al.]{
Andrea Verdini$^{1}$\thanks{E-mail: andrea.verdini@unifi.it},
R. Grappin$^{2}$,
O. Alexandrova$^{3}$,
L. Franci$^{4}$,
S. Landi$^{1}$,
\newauthor
L. Matteini$^{3}$,  
and E. Papini$^{1}$
\\
$^{1}$Dipartimento di Fisica e Astronomia, Universit\'a di Firenze, Firenze, IT\\
$^{2}$LPP, Ecole Polytechnique, Palaiseau, FR\\
$^{3}$LESIA, Observatoire de Paris, Universit\'e PSL, CNRS, Sorbonne Universit\'e, Univ. Paris Diderot, Sorbonne Paris Cit\'e, Meudon, FR\\
$^{4}$School of Physics and Astronomy, Queen Mary University of London, London, UK\\
}

\date{Accepted XXX. Received YYY; in original form ZZZ}

\pubyear{2019}

\begin{document}
\label{firstpage}
\pagerange{\pageref{firstpage}--\pageref{lastpage}}
\maketitle


\begin{abstract}
We analyse velocity fluctuations in the solar wind at magneto-fluid scales in two datasets, extracted from \textit{Wind} data in the period 2005-2015, that are characterised by strong or weak expansion.
Expansion affects measurements of anisotropy because it breaks axisymmetry around the mean magnetic field. 
Indeed, the small-scale three-dimensional local anisotropy of magnetic fluctuations ($\delta B$) as measured by structure functions ($SF_B$) is consistent with tube-like structures for strong expansion.
When passing to weak expansion, structures become ribbon-like because of the flattening of $SF_B$ along one of the two perpendicular directions. The power-law index that is consistent with a spectral slope $-5/3$ for strong expansion now becomes closer to $-3/2$. This index is also characteristic of velocity fluctuations in the solar wind.
We study velocity fluctuations ($\delta V$) to understand if the anisotropy of their structure functions ($SF_V$) also changes with the strength of expansion and if the difference with the magnetic spectral index is washed out once anisotropy is accounted for.
We find that $SF_V$ is generally flatter than $SF_B$. When expansion passes from strong to weak, a further flattening of the perpendicular $SF_V$ occurs and the small-scale anisotropy switches from tube-like to ribbon-like structures.
These two types of anisotropy, common to $SF_V$ and $SF_B$, are associated to distinct large-scale variance anisotropies of $\delta B$ in the strong- and weak-expansion datasets.
We conclude that $SF_V$ show anisotropic three-dimensional scaling similar to $SF_B$, with however systematic flatter scalings, reflecting the difference between global spectral slopes.
\end{abstract}

\section{Introduction}
     \label{sec:intro} 
The solar wind is a turbulent plasma that expands spherically in the heliosphere, with magnetic and velocity fluctuations having a power-law spectrum on several decades in frequency
(e.g. \citealt{Bruno_Carbone_2013} for a review). 
Above proton scales, where magnetohydrodynamics (MHD) is a good description of the plasma, the power-law index of magnetic fluctuations is on average $-5/3$, while velocity fluctuations have a flatter spectrum with an Iroshinikov-Kraichnan index, $-3/2$ (e.g. \citealt{Podesta_al_2007,Salem_al_2009,Tessein_al_2009}). The latter does not vary with properties of solar wind streams, while the magnetic index approaches a value $-3/2$ in strongly Alfv\'enic intervals \citep{Chen_al_2013} (Alfv\'enic intervals are those with a strong correlation between magnetic and velocity fluctuations).
The existence of different spectral indices shows that, one the one hand, homogenous MHD turbulence is a valid framework to interpret solar wind fluctuations, and, on the other hand, that some physical mechanisms beyond homogenous MHD turbulence may be needed to reproduce the observed properties. 
In fact, in homogenous incompressible MHD a cascaded quantity is associated to the existence of an ideally conserved quantity, namely the total (kinetic plus magnetic) energy. Since magnetic and kinetic energy are not conserved separately, there is no reason to expect different power laws for their spectra (we will come back on this point in the discussion), yet different spectral index are observed.

Theories generally assume the same spectral index for the two fields \citep[e.g.][]{GS95,B05,B06},
while numerical works either focus on the spectrum of the total energy or the magnetic energy (e.g. \citealt{Muller_al_2003,Beresnyak_Lazarian_2006,Mason_al_2006,Mason_al_2008,Grappin_Muller_2010,Perez_al_2012,Beresnyak_2015})
or, when studying magnetic and velocity spectra, they obtain different spectral indices that do not match the observed values (e.g. \citealt{Cho_Vishniac_2000,Milano_al_2001,Muller_Grappin_2005}). 
There are few numerical exceptions that report the spectral indices $-5/3$ and $-3/2$ for magnetic and velocity fluctuations, respectively: decaying two-dimensional (2D) hybrid simulations \citep{Franci_al_2015,Franci_al_2015b}, decaying 2D Hall-MHD simulations \citep{Papini_al_2019}, forced simulations of reduced MHD \citep{Boldyrev_al_2011}, and decaying simulations of full three-dimensional (3D) MHD without guide field \citep{Grappin_al_2016}.
Such simulations have different governing equations (full MHD, reduced MHD, Hall MHD, or fluid electrons and particle in cell protons), indicating that separate spectral indices can be obtained in a rather general context. However they also have different large scale dynamics (decaying or forcing) and different configurations (3D or 2D, with or without a mean field), which are not necessarily appropriate to describe solar wind turbulence that is expected to be 3D, decaying, and with a mean field of the order of the fluctuations.
Although a spectral relation between the total energy and residual energy (the difference between magnetic and kinetic energies) is a promising approach \citep{Grappin_al_1983,Muller_Grappin_2004,Muller_Grappin_2005,Grappin_al_2016}, with the residual energy originating from current sheets formed in the cascade process (e.g. \citealt{Matthaeus_Lamkin_1986}), there is currently no explanation for the difference in spectral indices of the magnetic and velocity fluctuations, 
Note, finally, that when the mean field is absent, the total energy spectrum can have a slope $-2$, $-5/3$, or $-3/2$ depending on initial condition or forcing \citep{Lee_al_2010,Krstulovic_al_2014}: even for a conserved quantity the spectral index may be non universal.

Solar wind turbulence is also anisotropic with respect to the mean-field direction.
Most of the works deal with magnetic field fluctuations, possibly because of the high cadence of the in-situ data and the association of strong currents with heating events \citep{Osman_al_2011,Osman_al_2012}.
Again, measurements of anisotropy with respect to the mean field, either calculated at large scale (global anisotropy) or at each scale (local anisotropy), show features that are characteristic of homogeneous MHD turbulence with some noticeable exceptions.

To be more specific, when global anisotropy is computed on solar wind data, magnetic fluctuations have a stronger power in the field-perpendicular wavevectors than in the field-parallel wavevectors, as expected for a plasma threaded by a mean field \citep{Montgomery_Turner_1981,Shebalin_al_1983,Grappin_1986, Muller_al_2003,Verdini_al_2015}.
However, in fast streams, magnetic fluctuations posses also the so-called slab component, with most of the energy residing in field-aligned wavevectors \citep{Matthaeus_al_1990,Bieber_al_1996,Dasso_al_2005,Weygand_al_2009,Weygand_al_2011}.  
This component has no stable counterpart in homogeneous MHD (\citealt{Ghosh_al_1998a,Ghosh_al_1998b} but see \citealt{Zank_al_2017} for an explanation based on nearly incompressible MHD). However, it can be easily explained as the result of the (wrong) assumption of axisymmetry around the mean field for structures that are instead axisymmetric around the radial direction. 
In fact, a radial symmetry emerges naturally when the non-linear dynamics is slower than the expansion of the solar wind \citep{Volk_Aplers_1973, Heinemann_1980,Grappin_al_1993,Verdini_Grappin_2016}, and is compatible with measurements at large scales \citep{Saur_Bieber_1999}. 
Whether the radial axis stops to rule the anisotropy at small scales depends on the tendency of turbulence to become strong, in analogy to the switch from weak to strong turbulence in homogenous MHD (e.g. \citealt{Verdini_Grappin_2012, Meyrand_al_2016}). For the solar wind, it is still unclear whether axisymmetry around the mean field is restored at proton scales \citep{Hamilton_al_2008, Narita_al_2010, Roberts_al_2017, Lacombe_al_2017}. 

When local anisotropy is computed, the spectral index is found to vary with the angle $\theta_{BV}$ between the mean field and the sampling direction, which is the radial direction of the wind flow. 
Most of the studies analysed magnetic field data in fast streams \citep{Horbury_al_2008,Podesta_2009, Wicks_al_2010, Wicks_al_2011,Wang_al_2014} and obtained spectral indices that pass from $-2$ to $-5/3$ with increasing $\theta_{BV}$ angle. These results are consistent with the anisotropy of homogenous strong turbulence, which is regulated by the critical balance between the linear Alfv\'en time and the non-linear eddy turnover time \citep{GS95}.
Exceptions are found in the works by \citet{Luo_Wu_2010} and \citet{Wang_al_2016}. In the former, the spectral index in the perpendicular direction was closer to $-3/2$, a value predicted when magnetic and velocity fluctuations progressively align at small scales \citep{B05,B06}. \citet{Wang_al_2016}, instead, found a weaker field-parallel index ($-1.75$) by requiring the magnetic field direction to be stable at large scales. However this result is not completely understood. The stability requirement alone seems not to be sufficient to return a weaker parallel spectral index \citep[see the analysis in][]{Gerick_al_2017}, although it limits the intermittency in the analysed intervals. Removal of intermittency from data yields flatter parallel spectra \citep{Wang_al_2014}, but the same procedure only affects perpendicular spectra in numerical simulations \citep{YangL_al_2017}.
 
\begin{figure}    
               \includegraphics[width=0.325\linewidth,clip=,trim={10cm 4.5cm 12cm 6cm}]{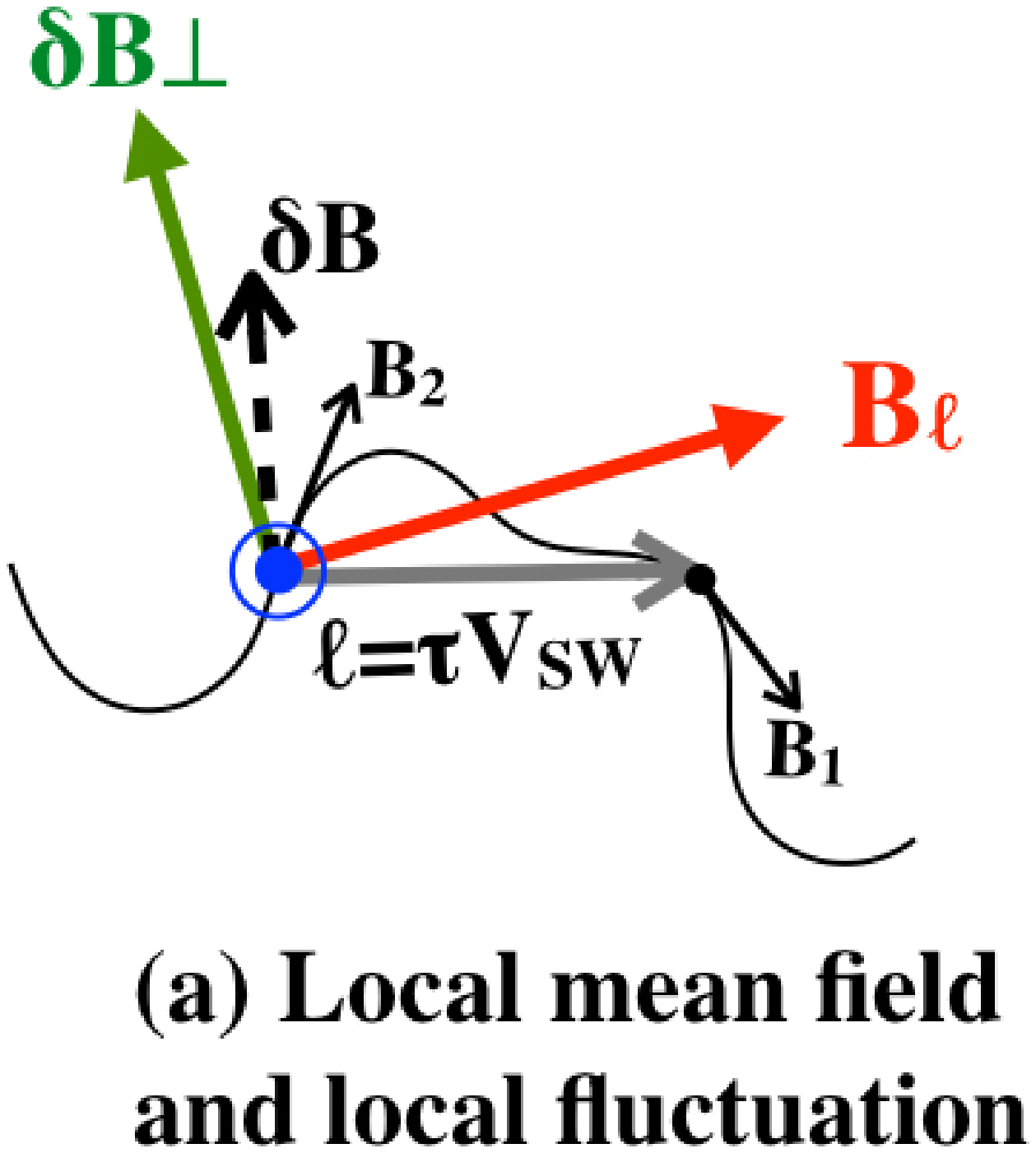}
               \includegraphics[width=0.325\linewidth,clip=,trim={8cm 3.5cm 13cm 6cm}]{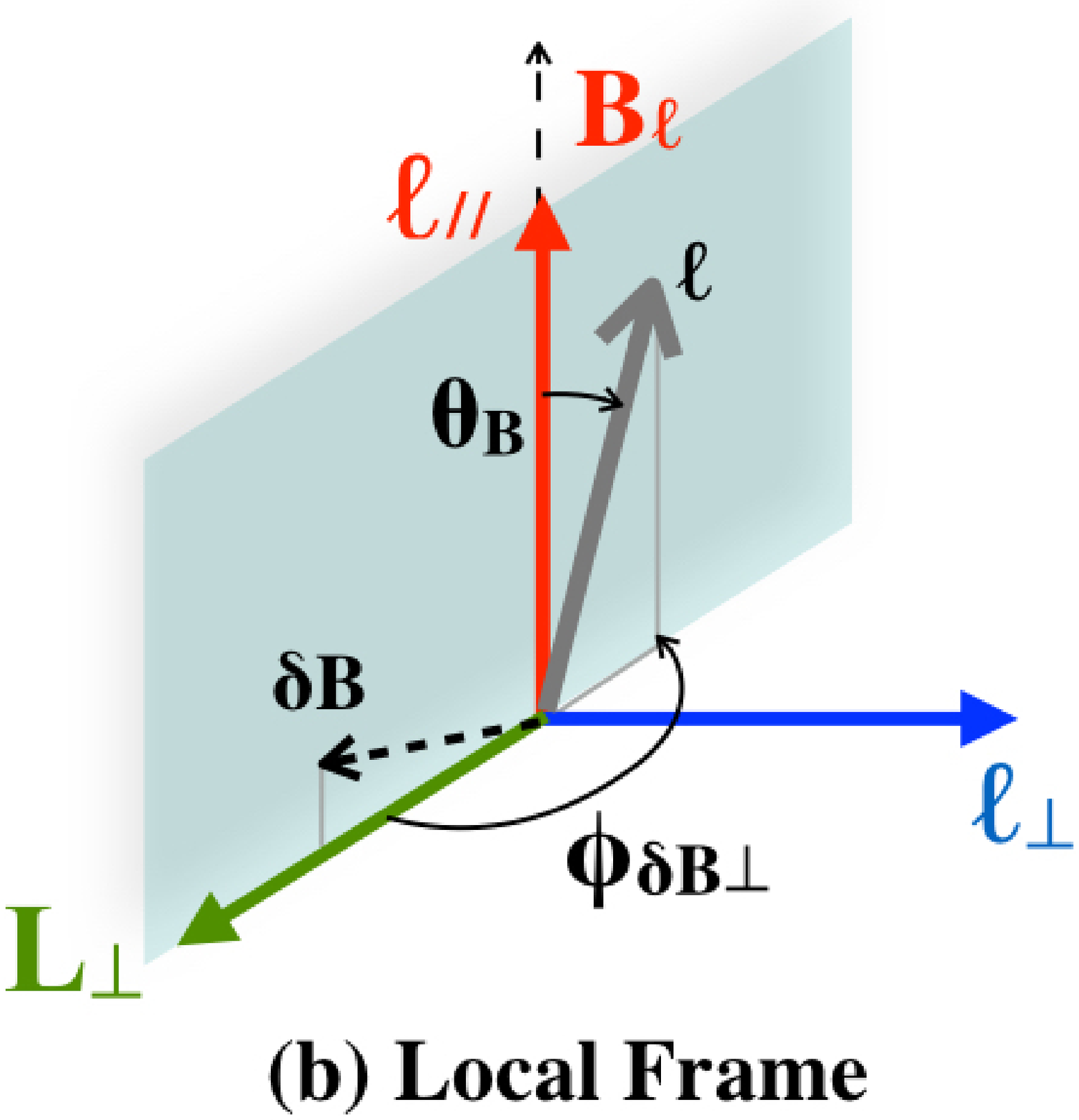}
               \includegraphics[width=0.325\linewidth,clip=,trim={10cm 6cm 11cm 8cm}]{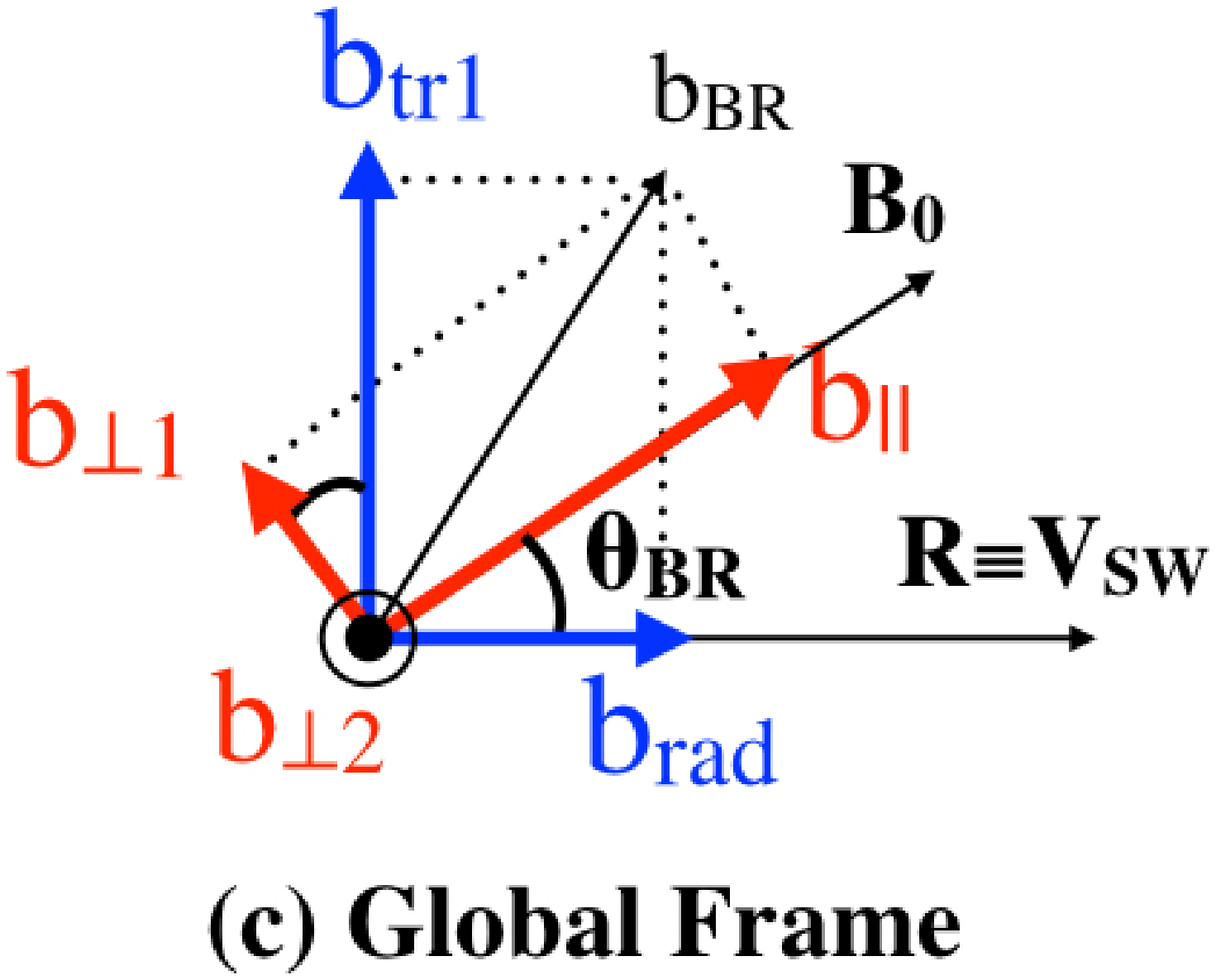}
              \caption{(a) From the measurements of magnetic fields at two different times, $\vect{B}_1$ and $\vect{B}_2$ (black arrows), separated by the lag $\tau$, one obtains the local mean field $\vect{B}_\ell$ (red arrow) and the local fluctuation $\delta \vect{B}$ (black dashed arrow). The displacement direction (green arrow) lies in the plane defined by $\delta B$ and $B_\ell$ and is perpendicular to the latter. The perpendicular direction (blue) complete the reference frame and is orthogonal to the plane of the figure. (b) The reference frame as obtained from the configuration on the left. The increment along the sampling direction is indicated with a grey arrow and its orientation in spherical coordinates is measured with the polar and azimuthal angles, $\theta_B,~\phi_{\delta B\bot}$, respectively. For purpose of illustration we choose the mean flow $\vect{V}_{sw}$ to be coplanar with $\vect{B}_1$ and $\vect{B}_2$ (the light-blue shaded plane), in the general case the vector $\vect{\ell}$ has a random orientation in the perpendicular plane where $\phi_{\delta\vect{B}\bot}$ is measured. (c) Relation between the radial-aligned and field-aligned reference frames (blue and red arrows, respectively). $b_{BR}$ is the projection of the fluctuation in the $BR$ plane containing the radial and large-scale mean-field directions that form an angle $\theta_{BR}$.
                      }
   \label{fig:frame}
   \end{figure}

Relaxation of axisymmetry in the measurements of local anisotropy (3D anisotropy) revealed that structures have their largest dimension in the field-parallel direction and are axisymmetric around this axis only at small scales \citep{Chen_al_2012}. At large scales, turbulent eddies have their largest dimension in the displacement direction, that is, the direction perpendicular to the mean field and with a component along the fluctuation direction (the proper perpendicular direction is perpendicular to both the mean field and the fluctuation, see Figure~\ref{fig:frame}a,b).
Numerical simulations of MHD turbulence with the Expanding Box Model (EBM, \citealt{Velli_al_1992,Grappin_al_1993,Grappin_Velli_1996}), allowed interpreting this unusual large-scale anisotropy as a consequence of the spherical expansion of the solar wind, which introduces a radial symmetry in the amplitude of magnetic field fluctuations, with radial fluctuations being less energetic than those transverse to the radial (see \citealt{Dong_al_2014}  for 3D simulations and section~\ref{sec:method} for an explanation based on the conservation of the magnetic flux). 

In addition, there are indications that expansion can alter also the small-scale anisotropy. 
In fact, the above simulations \citep{Verdini_Grappin_2015} also showed that the spectral indices in the perpendicular and displacement directions are the same when data are sampled in the radial direction, in agreement with observations \citep{Chen_al_2012}, while they are different when sampling in directions transverse to the radial. These numerical results were partially confirmed by a two-spacecraft analysis \citep{Vech_Chen_2016}.
Following these findings, \citet{Verdini_al_2018} computed the 3D local anisotropy of magnetic fluctuations in two datasets in which the effects of expansion are expected to be large and weak, respectively.
For strong expansion they recovered the same anisotropy as in \citet{Chen_al_2012}, in agreement with axisymmetry at small scales, as predicted by the critical balance \citep{GS95}. Instead, for weak expansion, they obtained different spectral indices in the perpendicular and displacement directions, that is, non-axisymmetric structures similar to ribbons, an anisotropy predicted by \citet{B05,B06}.

Velocity fluctuations have not been studied in such detail, possibly because of the lower resolution of plasma data, although vortical structures are ubiquitous in the solar wind \citep{Perrone_al_2016,Perrone_al_2017} and small-scale vorticity enhancements are shown to be co-spatial with preferential perpendicular heating of protons in 2D hybrid simulations \citep{Franci_al_2016b}.
As a result, their anisotropy is less constrained. 
First, there is no measurement of the 3D anisotropy of velocity fluctuations. 
Second, the measurements of axisymmetric anisotropy yield contradictory results.
On the one hand, by analysing a single fast stream in the ecliptic, \citet{Wicks_al_2011} found angle-dependent power-law indices, the index decreasing monotonically from $-2$ to $-3/2$ when $\theta_{BV}$ passes from $0^o$ to $90^o$. On the other hand, by averaging seven fast streams including the previous one, \citet{Wang_al_2014} found that the index of velocity fluctuations was consistent with an angle-independent value of $-3/2$, although in some particular streams the field-parallel direction had an index $-2$. They also suggested that differences in velocity and magnetic anisotropy arise from intermittency. In fact, when intermittency was removed from the data they obtained similar and angle-independent indices for the magnetic and velocity fluctuations. 
This is at odds with earlier analysis of solar wind data and with recent numerical simulations. 
\citet{Salem_al_2009} removed intermittency from \textit{Wind} data and still obtained spectral indices of $-5/3$ and $-3/2$ for the magnetic and velocity fluctuations, respectively (although they did not consider the anisotropy with respect to the mean field). Upon removal of intermittency in direct numerical simulation of compressible MHD, \citet{YangL_al_2017} found only small variations of the spectral index anisotropy, with indices being always the larger in the field-parallel direction. 

In this work we extend the 3D analysis of local anisotropy of magnetic fluctuations \citep{Chen_al_2012,Verdini_al_2018} to velocity fluctuations, by analysing separately intervals with weak and strong expansion.
In section~\ref{sec:method} we briefly describe the method used to construct the datasets and to analyse the anisotropy via second order structure functions. In section~\ref{sec:data} we complement the characterization of the two datasets given in \citet{Verdini_al_2018}. In section~\ref{sec:axis} we present the results on the local anisotropy of magnetic and velocity fluctuations under the assumption of axisymmetry. We then show the 3D anisotropy of velocity fluctuations in section~\ref{sec:3Danis}. In section~\ref{sec:discussion} we summarise and discuss the results.

\section{Data and Method of Analysis} 
      \label{sec:method}      
We briefly describe the data and the method used in the analysis, more details can be found in \citet{Verdini_al_2018}.
We use data at 1AU from instruments on \textit{Wind} spacecraft in the period 2005-2015: magnetic field data at $3s$ resolution from MFI instrument \citep{Lepping_al_1995} and onboard ion moments at $3s$ resolution from 3DP/PESA-L \citep{Lin_al_1995}.
To separate intervals with weak and strong expansion we compute the ratio
\begin{equation}
E=\frac{b_{tr}^2}{b_{rad}^2}=\left.\frac{b_Y^2 + b_Z^2}{b_X^2}\right|_{{\cal T}=2h},
\label{crit}
\end{equation}
where $(X,Y,Z)$ are the GSE coordinates (with X aligned with the radial direction), the subscripts $rad$ and $tr$ refer to the radial and transverse-to-the-radial components, and the fluctuations $b_{X,Y,Z}$ are obtained by subtracting a running average with a window of duration ${\cal T}=2h$ from the original signal: $\vect{b}=\vect{B}-\langle \vect{B}\rangle_{\cal T}$.
The two datasets contain intervals of at least 5h that satisfy at each time the following criteria:
The two datasets contain intervals of minimal duration that satisfy continuously for at least 5h the following criteria:
\begin{eqnarray}
2<E<10\;\; \mbox{strong-expansion dataset} \label{crit1},\\
0<E<2\;\;  \mbox{weak-expansion dataset} \label{crit2}.
\end{eqnarray}
Continuously means that the criterion must be satisfied at each time belonging to the interval, and not only on average in the interval, although we allow for out-of-bounds values of $E$ on a duration of 1 minute (see figure 1 in \citealt{Verdini_al_2018}).

The relation between $E$ and expansion is now briefly explained, more details can be found in \citet{Grappin_Velli_1996, Dong_al_2014, Verdini_Grappin_2015,MontagudCamps_al_2018}. For a spherically expanding flow, the conservation of magnetic flux imposes a weaker decay of transverse components of magnetic fluctuations compared to the radial one, $b_{tr}\propto1/R$ and $b_{rad}\propto1/R^2$. Assuming component isotropy close to the Sun ($b_Y\sim b_Z\sim b_X$ and so $E=2$), one finds stronger transverse fluctuations at 1AU, with $E>2$. Note, however, that velocity fluctuations must satisfy the conservation of angular momentum, $u_{tr}\propto 1/R$ and $u_{rad}=const$, resulting in the opposite behaviour, that is, a faster decay of transverse components. Any form of coupling between velocity and magnetic fluctuations has the effect of smearing out the anisotropy evaluated by $E$. 
For a strong linear Alfv\'enic coupling, all the components of velocity and magnetic fluctuations decay as $1/\sqrt{R}$, thus maintaining the value of $E$ close to the Sun \citep[e.g.][]{Dong_al_2014}. Instead, when the nonlinear coupling is strong, a component anisotropy with respect to the mean magnetic field direction develops \citep[e.g.][]{Oughton_al_2016}.
At large scales (of the order of few hours in the solar wind) one expects the expansion timescale to be shorter than the nonlinear timescale and the Alfv\'en timescale, so that $E$ is ruled by the conservation of magnetic flux. 
This is confirmed by numerical simulations of MHD turbulence in 3D with a mean magnetic field of the order of the fluctuations. Without expansion and with an oblique mean field, $E\approx 2 $ at all times, while in presence of expansion $E$ increases monotonically with distance and does not depend much on the scale at which is computed \citep{Verdini_Grappin_2015}.
It should be noticed that requiring $E>2$ to isolate intervals with strong expansion is only an approximate criterion, since the divergence-less condition of magnetic fluctuations contributes to such inequality as much as expansion in single spacecraft measurements \citep{Vech_Chen_2016}.

Although we will compute the local anisotropy of velocity structure functions, we define the local reference frame with respect to the local mean magnetic field and the local magnetic fluctuation as in \citet{Chen_al_2012}. 
For each pair of magnetic field $\vect{B}_1=\vect{B}(t)$, $\vect{B}_2=\vect{B}(t+\tau)$ separated by a time lag $\tau$, the fluctuation is defined as
\be
\delta \vect{B}=\vect{B}_1-\vect{B}_2,
\label{deltaB}
\ee 
while the local mean field is given by 
\be
\vect{B}_l=1/2(\vect{B}_1+\vect{B}_2).
\label{Blocal}
\ee 
We choose the $z$ axis along the mean field, the $x$ axis along the local perpendicular displacement direction, 
\be
\delta\vect{B}_\bot\propto\vect{B}_l\times[\delta\vect{B}\times\vect{B}_l],
\label{deltaBperp}
\ee
and the $y$ axis, the perpendicular direction, is orthogonal to both the fluctuation and the mean field (see Figure~\ref{fig:frame}a).
We use a spherical polar coordinate system in which the radial vector $\vect{\ell}$ coincides with the solar wind flow direction, i.e. the sampling direction, and use the polar $\theta_B$ and azimuthal $\phi_{\delta B\bot}$ angles to measure its orientation with respect to the mean-field and the displacement directions, respectively (see Figure~\ref{fig:frame}b).

For each pair of points, the square of the velocity fluctuation is binned in this 3D coordinate system, and the 
velocity structure function is defined as
\be
SF_i(\ell,\theta_B,\phi_{\delta B\bot})=\langle\delta \vect{V}^2\rangle_i=\langle|\vect{V}_1-\vect{V}_2|^2\rangle_i,
\label{base}
\ee
where we have indicated with $\langle...\rangle_i$ an average on all increments computed in the interval $i$.
We use 66 logarithmically spaced increments to measure the power level in the range 
$10^{-4}~\mathrm{Mm^{-1}}<k<1~\mathrm{Mm^{-1}}$, where $k=1/\ell$ is the wavenumber obtained from the increment $\vect{\ell}=\tau \vect{V}_{SW}$. 
The sampling direction is given by the solar wind speed, $\vect{V}_{SW}$, which is the average of the first moment of the ion distribution computed in each interval, $\vect{V}_{SW}=\langle\vect{V}\rangle_i$ (using a local definition $\vect{V}_{SW}(\ell)=1/2(\vect{V}_1+\vect{V}_2)$ does not change the results).
For the polar and azimuthal angles we use $5^o$ bins to cover one quadrant only (any angle greater than $90^o$ is reflected below $90^o$).

To obtain a $SF$ for a given dataset, before averaging among intervals we normalise each structure function, $SF_i(\ell,\theta_B,\phi_{\delta B\bot})$, by the energy of velocity fluctuations at a scale $\ell^*=100~\mathrm{Mm}$, which is in the middle of the power-law range of their spectrum. The energy is obtained by averaging over angles
\be
S_i(\ell^*)=\sum_{\theta_B,\phi_{\delta B\bot}} w_i SF_i 
\label{energy}
\ee
with weights given by 
\be
w_i=N_i(\ell,\theta_B,\phi_{\delta B\bot})/N_i(\ell)
\ee
where $N_i(\ell)=\sum_{\theta_B,\phi_{\delta B\bot}}N_i(\ell,\theta_B,\phi_{\delta B\bot})$.
The average among intervals is weighted with the relative count in each bin, 
so that the structure function is given by, 
\be
SF(\ell,\theta_B,\phi_{\delta B\bot})=\langle w_i SF_i / S_i(\ell^*)\rangle_i.
\ee
with weights
\be
w_i=N_i(\ell,\theta_B,\phi_{\delta B\bot})/N(\ell,\theta_B,\phi_{\delta B\bot}).
\ee

From the recorded values of $SF_i$ we can also compute the axisymmetric $SF$ by averaging along the azimuthal angle $\phi_{\delta B\bot}$.
The axisymmetric $SF$ will be used for comparison with previous works and is obtained as
\be
SF(\ell,\theta_B)=\left\langle\sum_{\phi_{\delta B\bot}} \left[  w_iSF_{i}/S_i(\ell^*)\right] \right\rangle_i,
\label{axis}
\ee
with weights given by
\be
w_i=N_i(\ell,\theta_B,\phi_{\delta B\bot})/N(\ell,\theta_B),
\label{eq:waxi}
\ee
in which we have defined $N(\ell,\theta_B)=\sum_{i,\phi_{\delta B\bot}}N_i $.

We will also compute the \textit{raw} axisymmetric $SF$ that is obtained in a similar way but without applying any normalisation to the $SF_i$ that belong to the same dataset, that is,
\be
SF(\ell,\theta_B)=\left\langle\sum_{\phi_{\delta B\bot}}  \left[w_i SF_{i}  \right]\right\rangle_i.
\label{axis-raw}
\ee
with $w_i$ in eq.~\ref{eq:waxi}. This raw $SF$ will be used to evaluate the signal to noise ratio in velocity structure functions.
We will also show the axisymmetric $SF$ for magnetic fluctuations, $SF_B$, which is obtained by collecting the power $\langle|\vect{B}_1-\vect{B}_2|^2\rangle_i$ in eq.~\ref{base}. This power is then used in eq.~\ref{energy} to compute the magnetic energy for the (eventual) normalization.

\begin{figure}    
	      \includegraphics[height=0.600\linewidth,clip=,trim={0.7cm 0.cm 0.4cm 0cm}]{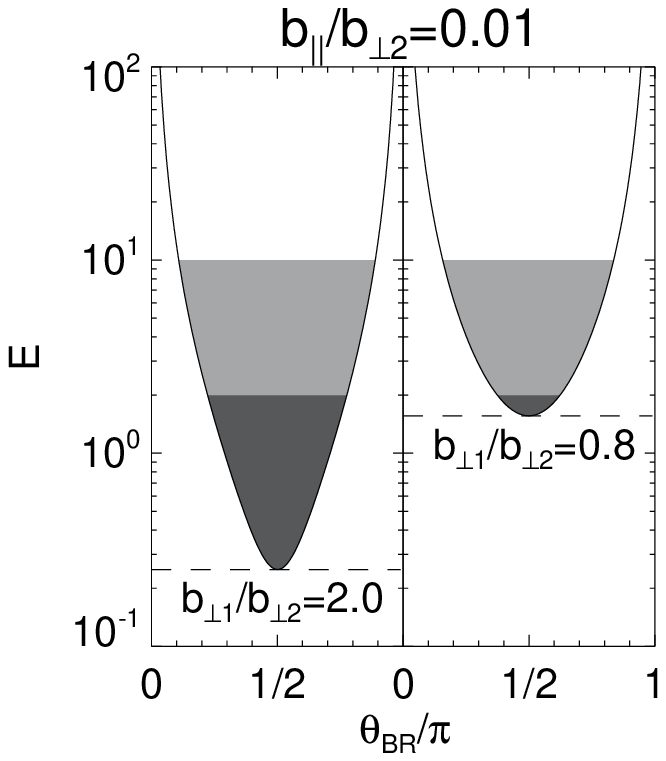}
	      \includegraphics[height=0.600\linewidth,clip=,trim={2.1cm 0.cm 0.4cm 0cm}]{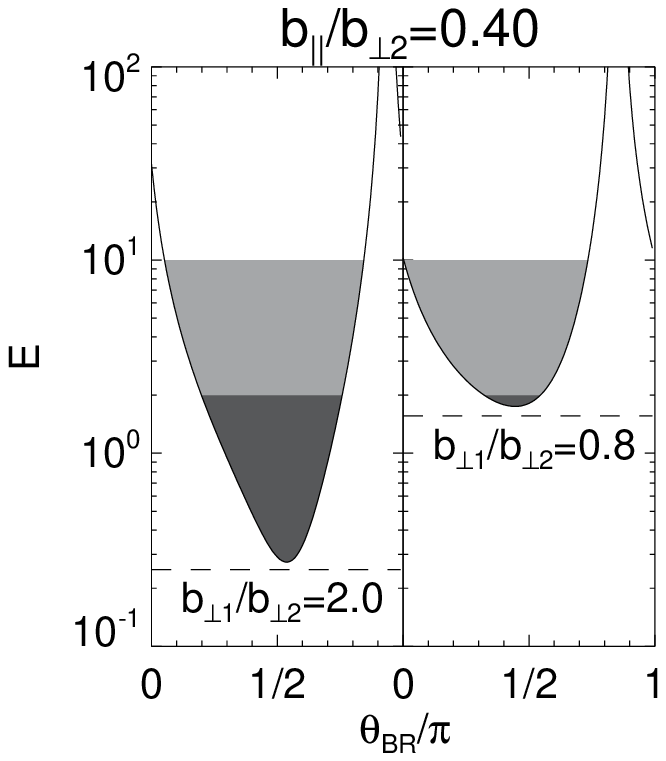}      
             \caption{The ratio $E$ as a function of $\theta_{BR}$ (see eq.~\ref{eq:bias}) with $b_{\bot1}/b_{\bot2}=2,~0.8$. The horizontal dashed line is the value $b^2_{\bot2}/b^2_{\bot1}$ that corresponds to the minimum of $E$ at $\theta_{BR}=\pi/2$ only for small $b_\|/b_{\bot2}$ (0.01 in the left panel). When $b_\|/b_{\bot2}$ increases (0.4 in the right panel) the function becomes asymmetric, the minimum shifts to larger values and angles, while the asymptotes moves to smaller angles. The light and dark grey areas correspond to the values of $E$ that determine the strong- and weak-expansion datasets, respectively.}
   \label{fig1d}
   \end{figure}

\section{Dataset Properties} 
 \label{sec:data} 
   \begin{figure}    
	      \includegraphics[width=0.980\linewidth,clip=,trim={0.8cm 0.cm 0.2cm 0cm}]{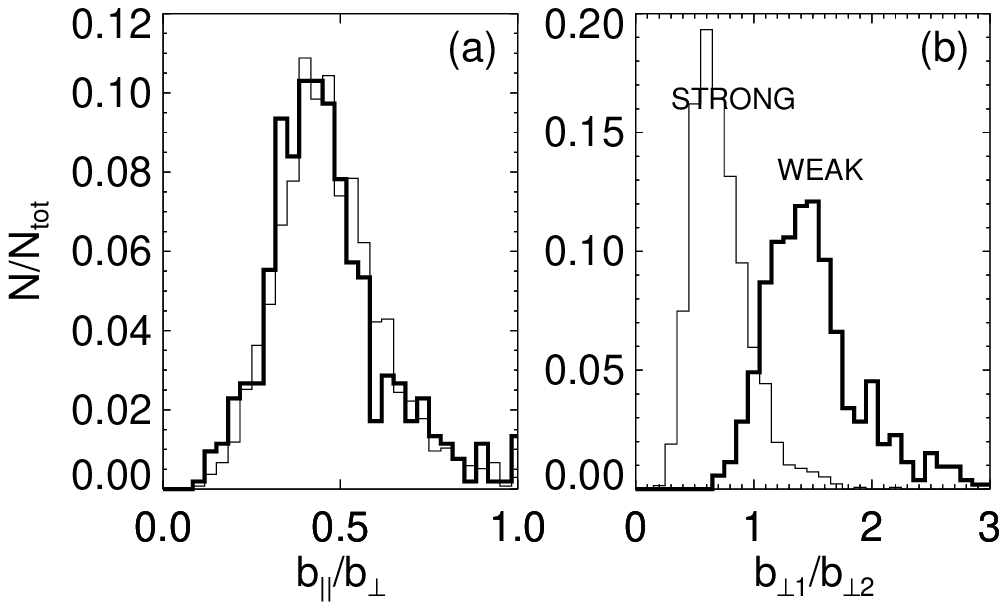}\\
	      \includegraphics[width=0.980\linewidth,clip=,trim={0.8cm 0.cm 0.2cm 1.4cm}]{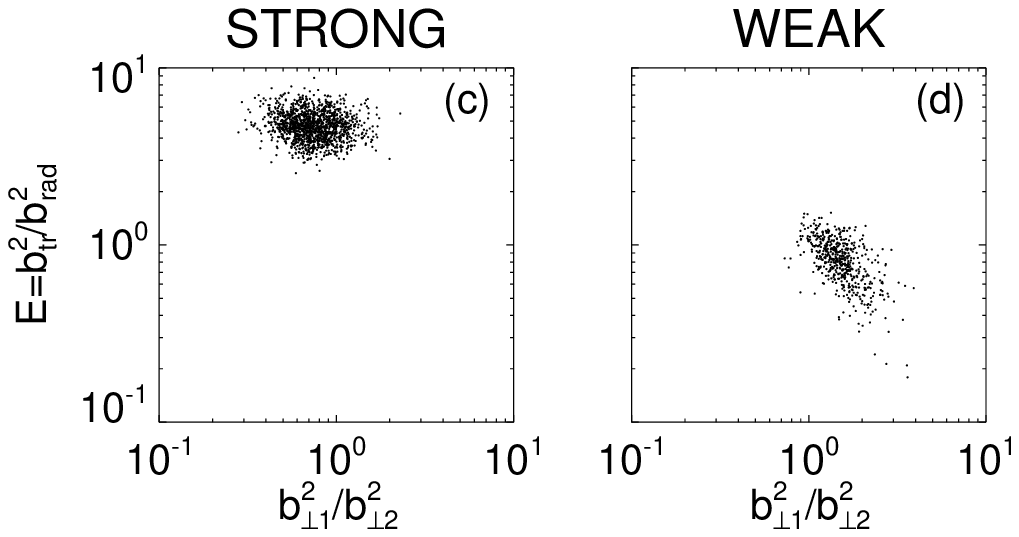}
              \caption{Top panels: distribution in the strong- and weak-expansion datasets (thin and thick lines, respectively) of the ratio between the amplitude of field-parallel and field-perpendicular magnetic fluctuations, $b_{\|}/b_{\bot}$ (left) and of the amplitudes of magnetic fluctuations in the two field-perpendicular components, $b_{\bot1}/b_{\bot2}$ (right). Bottom panels: scatterplot of $b^2_{\bot1}/b^2_{\bot2}$ versus $E=b^2_{tr}/b^2_{rad}$ for the strong- and weak-expansion datasets in the bottom left and bottom right panels, respectively.
              }
   \label{fig1}
   \end{figure}
\begin{figure*}    
               \includegraphics[width=0.980\linewidth,clip=,trim={0cm 0.cm 0cm 0cm}]{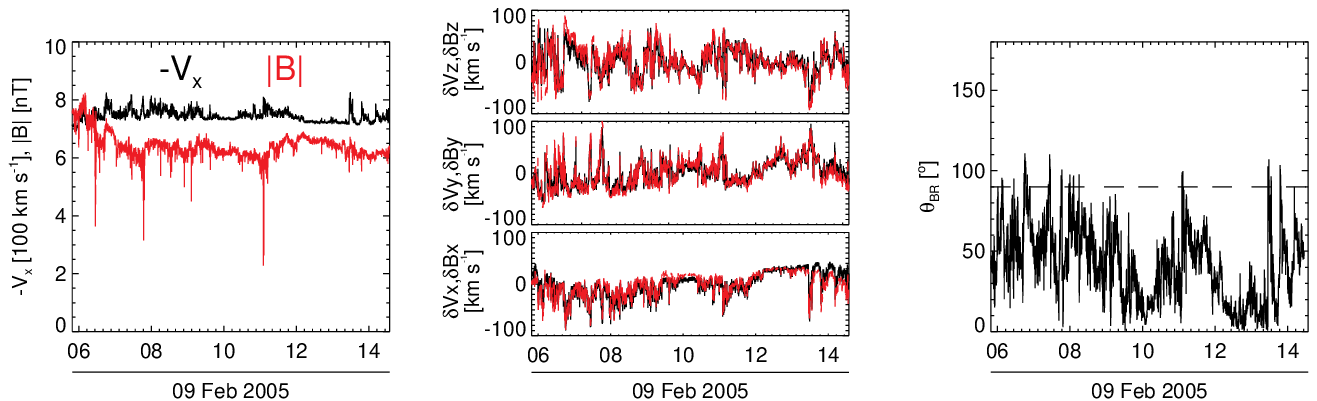}\\               	
               \includegraphics[width=0.980\linewidth,clip=,trim={0cm 0.cm 0cm 0cm}]{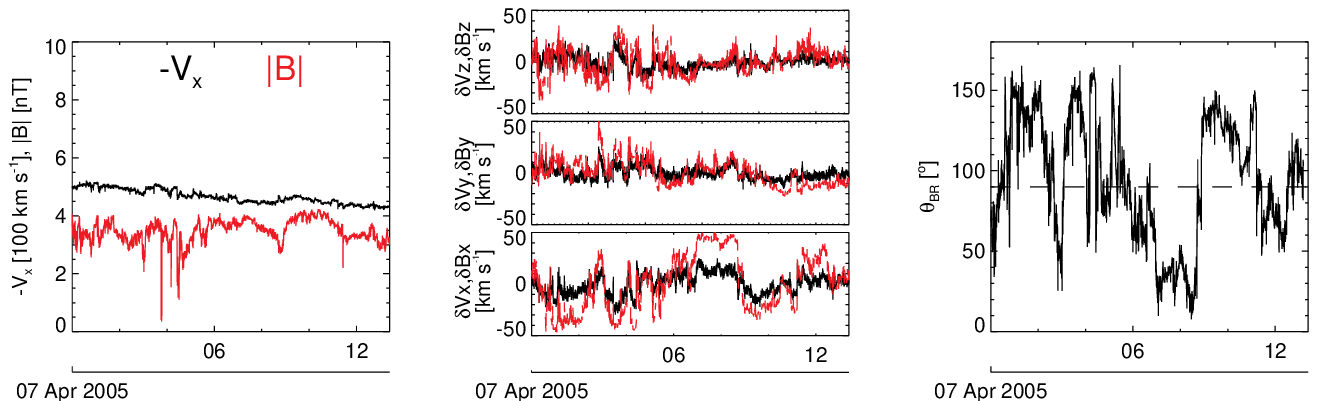}            
              \caption{Time series of two intervals representative of the strong-expansion dataset (top panels) and weak expansion dataset (bottom panels). Left column: solar wind radial velocity, $V_R=-V_x$ (black), and magnetic field intensity, $|B|$ (red). Central column: components of velocity and magnetic fluctuations, $\delta V_{X,Y,Z}$ and $\delta B_{X,Y,Z}$ in black solid and red dashed lines, respectively. Right column: magnetic field angle with respect to the radial direction, $\theta_{BR}$.
                      }
   \label{fig1a}
\end{figure*}

\subsection{Implication of the selection criterion}
The ratio $E$, used to separate weak-expansion and strong-expansion intervals, introduces a bias on the mean magnetic field direction and on the relative magnitude of the components of magnetic fluctuations (component anisotropy or variance anisotropy).
To see how, let us rewrite the ratio $E=b_{tr}^2/b_{rad}^2$ in the reference frame attached to the mean magnetic field direction, assumed to form an angle $\theta_{BR}$ with the radial direction, as represented in Figure~\ref{fig:frame}c. Indicating the components of fluctuations in this reference frame with $(b_{\|},~b_{\bot1},~b_{\bot2})$, with $b_{\bot1}$ lying in the $BR$ plane, we have
\begin{eqnarray}
E=\frac{b_{tr1}^2+b_{\bot2}^2}{b_{rad}^2}=\frac{\left(b_\| \sin\theta_{BR} + b_{\bot1} \cos\theta_{BR}\right)^2+b_{\bot2}^2}{\left(b_\| \cos\theta_{BR} + b_{\bot1} \sin\theta_{BR}\right)^2}\label{eq:bias}\\
\approx\frac{(b_{\bot2}/b_{\bot1})^2 + \cos^2\theta_{BR}}{\sin^2\theta_{BR}},\label{eq:bias1}
\end{eqnarray}
where in the last equality we have assumed $b_\|\ll b_{\bot1},b_{\bot2}$, that is, a weak magnetic compressibility.
\cha{It is worth mentioning that this strong inequality is used here as a simplification assumption, but it does not alter the final results (see below). In the solar wind one rather finds $b_\| \lesssim 1/2 b_{\bot}$, see for example fig.~\ref{fig1}a. We recall that incompressible MHD turbulence can maintain variance isotropy ($b_\|\sim b_{\bot}$), while  $b_\| < b_{\bot}$ is an asymptotic stated of decaying  weakly compressible MHD turbulence \citep[see][]{Matthaeus_al_1996,Oughton_al_2016}}.
The function is periodic with period $\pi$ and its form (eq.~\ref{eq:bias1}) is shown in the left panel of Figure~\ref{fig1d} for two values of the parameter $b_{\bot1}/b_{\bot2}=2,~0.8$. This parameter also set the minimum, $E_{min}=b^2_{\bot2}/b^2_{\bot1}$, which is always located at $\theta_{BR}=\pi/2$: the larger the ratio, the smaller is the minimum. With the chosen parameters that are indicated in the figure, $E_{min}=0.25,1.56$.

The dark shaded area corresponds to the selection criterion of the weak-expansion dataset, $0<E<2$. It is clear that the distribution of $\theta_{BR}$ is peaked at $\pi/2$ for a given value of the parameter $b_{\bot1}/b_{\bot2}$, but not all values are allowed. If one decreases further the ratio, for $b_{\bot1}/b_{\bot2}<1/\sqrt{2}$ the minimum value of $E$ is larger than 2 and no interval can satisfy the weak-expansion constraint.
When the field-aligned fluctuations are not negligible (right panel for which $b_\|/b_{\bot2}=0.4$), the function becomes asymmetric, the minimum of $E$ increases and shifts to larger angles, but the above considerations remain valid as far $b_{\|}/b_{\bot2}\lesssim1$.
The grey shaded area corresponds to the selection criterion of the strong-expansion dataset, $2<E<10$. 
Again the value of $E_{min}$ puts an upper limit to $b_{\bot1}/b_{\bot2}<1/\sqrt{10}$, for larger value the dataset is empty. Requiring $E<10$ also has the effect of excluding intervals with mean-field almost aligned to the radial direction: the lower the ratio $b_{\bot1}/b_{\bot2}$, the larger the angles that are omitted. However, as the field-parallel fluctuations are non-negligible, intervals with radial mean field direction are no more excluded and smaller ratios of $b_{\bot1}/b_{\bot2}$ are allowed (right panel).

The above constraints on the field-perpendicular components can be seen in the distributions shown in Figure~\ref{fig1}.
In the top left panel, one can see that the condition $b_\|/b_\bot<1$ is generally satisfied and the distributions are very similar in both datasets. On the contrary, the value $b_{\bot1}/b_{\bot2}=1$ roughly separates the distributions in the top right panel, the ratio being generally smaller for strong expansion and larger for weak expansion, with peak values at 0.5 and 1.5, respectively.
We also show a scatterplot of $E$ versus the ratio $b^2_{\bot1}/b^2_{\bot2}$ for strong and weak expansion in the bottom left and right panels, respectively. While for strong expansion no correlation can be seen, for the weak expansion a clear anti-correlation appears because of the relation $E_{min}=b^2_{\bot2}/b^2_{\bot1}$ which bounds $E$ from below. 
In other words, values $b_{\bot1}/b_{\bot2}>1$ for weak expansion are expected as a consequence of the selection criterion, while for strong expansion the selection criterion puts no constraint on the perpendicular component anisotropy. However, its distribution spans basically values $b_{\bot1}/b_{\bot2}<1$. A ratio smaller than one can be understood as a consequence of the divergence-less condition for the magnetic field. In fact, the fluctuation $b_{\bot1}$ is perpendicular to $B_0$ but has a non-vanishing projection on the wavevector $k$ along the radial sampling direction. On the contrary, $b_{\bot2}$ is perpendicular to both $B_0$ and $k$. If the two components have the same power and spectral index ($\alpha$), for a an oblique mean field and a sampling along the radial direction the ratio $b^2_{\bot1}/b^2_{\bot2}=1/\alpha$ at all scales \citep{Saur_Bieber_1999}.

To summarize, the combination of three factors determine the ordering in the two datasets of the amplitudes of the components of magnetic fluctuations as seen in the field-aligned reference frame (variance or component anisotropy).
These factors are: the selection criterion, the divergence-less constraint of magnetic fluctuation, and the smaller amplitude of field-aligned fluctuations compared to field perpendicular fluctuations (the latter is also a consequence of the divergence-less constraint if energy resides mostly in field-perpendicular wavevectors).
For the strong expansion dataset one has $b_\|<b_{\bot1}<b_{\bot2}$, while for the weak expansion dataset the ordering of the perpendicular components is reversed, $b_\|<b_{\bot2}<b_{\bot1}$. 
Note that on the one hand, the selection criterion for weak expansion also forces the mean magnetic field to be preferentially perpendicular to the radial direction, and since fluctuations in the $BR$ plane are large, the magnetic field direction is expected to vary substantially (see below). On the other hand, the mean-field direction is much less  constrained for the strong expansion dataset and can take basically any value.

\subsection{Dataset properties, two representative intervals}
We now describe the properties of the two datasets, first by inspecting one interval per dataset and then by looking at the distributions of the properties of intervals in each dataset.
The time series of (i) the radial velocity, the magnetic field intensity (left panels), (ii) the three components of velocity and magnetic fluctuations in Alfv\'en units (black and red lines respectively, central panel), and (iii) the angle between the magnetic field and the flow direction, $\theta_{BR}$  (right panels), are plotted in Figure~\ref{fig1a} for two  intervals representative of the strong-expansion dataset (top) and of the weak-expansion dataset (bottom).

The strong-expansion interval is a fast stream of average radial speed of $V_R=-V_x\approx700~\mathrm{km~s^{-1}}$, on top of which several jets of the order of $50~\mathrm{km~s^{-1}}$ are visible (top left panel). Such velocity enhancements are (anti) correlated to the variations of the magnetic intensity (red line) and are related to the Alv\'enic nature of this interval. 
In fact, as can be seen in the central top panel, velocity and magnetic fluctuations are strongly correlated.  
Note that the $X$ and $Y$ components of the fluctuations are ``one-sided'', that is, asymmetric with respect to zero, a characteristic of Alfv\'enic fluctuations with constant $|B|$ \citep{Gosling_al_2009,Matteini_al_2014}, although here $|B|$ is not perfectly constant.
The amplitude of fluctuations is larger in the $Z$ component and about the same in the $X$ and $Y$ components, reflecting the selection criterion $E=b_{tr}/b_{rad}> 2$.
Finally, the magnetic field is on average aligned with the Parker spiral (top right panel), it has almost no change in its polarity with variations in the angle $\theta_{BR}$ being generally smaller than $45^o$ (only in few cases variations reach $90^o$ and are associated to the jets in the radial velocity).

The weak-expansion interval has a moderate wind speed of about $400~\mathrm{km~s^{-1}}$, with small radial velocity fluctuations, and a weaker and more variable magnetic intensity (bottom left panel). 
Velocity and magnetic fluctuations are about a factor 2 smaller than in the strong-expansion interval (compare the central panels), they are only weakly correlated and are about symmetric with respect to zero. 
Magnetic fluctuations (red lines) have larger amplitudes in the $X$ component, again reflecting the selection criterion $E=b_{tr}/b_{rad} < 2$. 
The magnetic field is now on average perpendicular to the radial direction (the dashed line in the bottom right panel), although the instantaneous direction varies largely, with fluctuations of the order of $90^o$ associated to a change in magnetic field polarity and intensity.
 \begin{figure*}    
      \includegraphics[width=0.980\linewidth,clip=,trim={0cm 0.cm 0cm 0cm}]{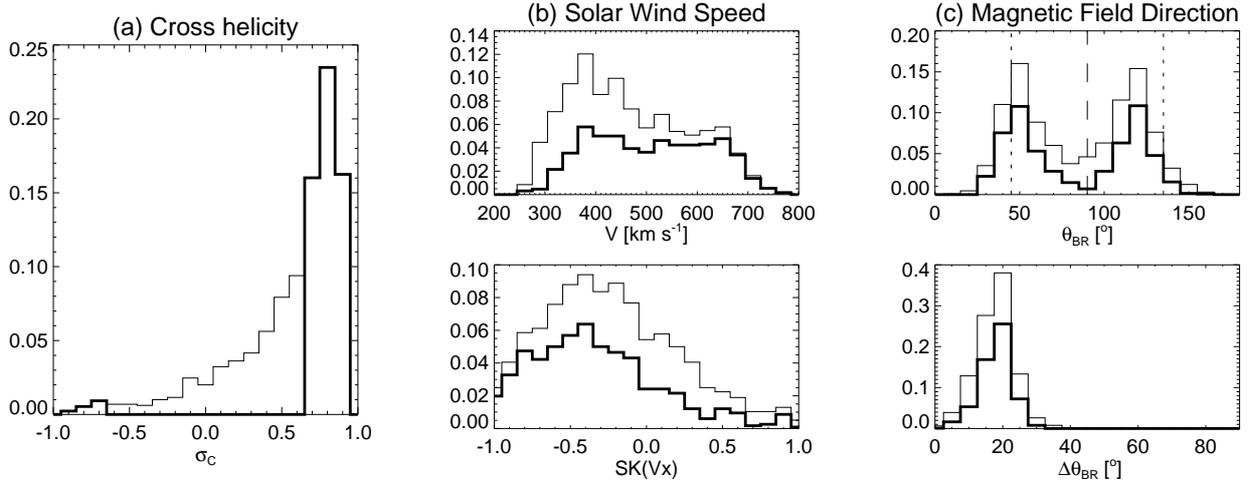}              
\caption{Strong-expansion dataset. Distribution of the normalized cross-helicity, $\sigma_c$, calculated with fluctuations in the frequency band $f\in[4.6,9.2]10^{-4}~\mathrm{Hz}$ (left panel), the mean solar wind speed and the skewness of its radial component, $V$ and $SK(V_x)$ (top and bottom central panels, respectively), the average angle between the magnetic field and the radial direction along with its standard deviation, $\theta_{BR}$ and $\Delta\theta_{BR}$ (top and bottom right panels, respectively). Thick histograms refer to a subsample with $|\sigma_c|\ge0.7$.
}
   \label{fig1b}
   \end{figure*}

\begin{figure*}    
	      \includegraphics[width=0.980\linewidth,clip=,trim={0cm 0.cm 0cm 0cm}]{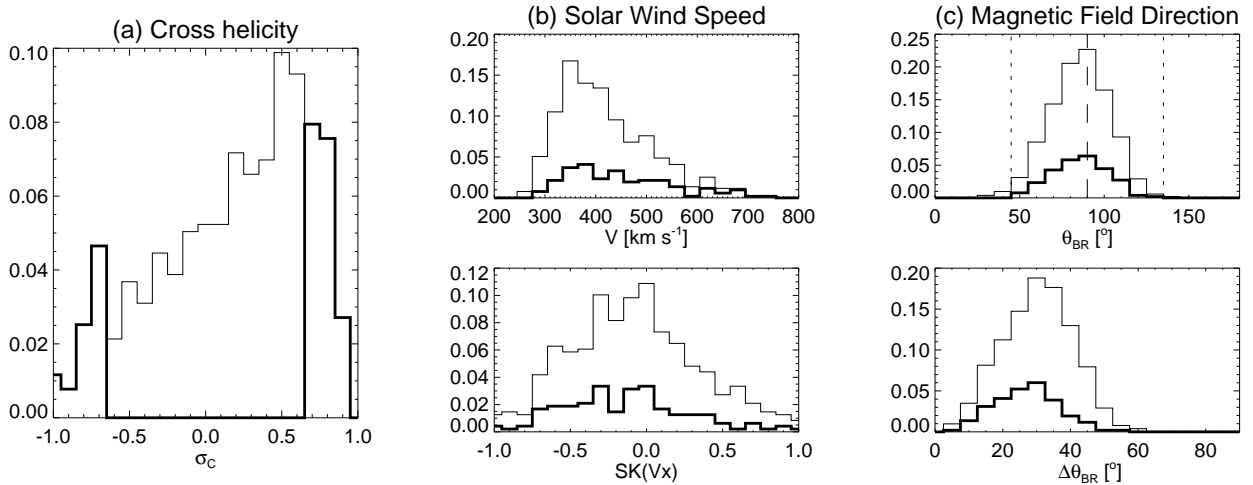}
              \caption{Weak-expansion dataset. Same as in Figure~\ref{fig1b} with the thick histograms referring to a subsample with $|\sigma_c|\ge0.7$}
   \label{fig1c}
   \end{figure*}

\subsection{Distribution of properties in intervals}
We finally show how properties related to those just commented above are distributed in the two datasets. 
For the strong-expansion dataset we plot in Figure~\ref{fig1b} the distributions of: (a) the average cross helicity $\sigma_c=-2\vect{u}\cdot\vect{b}/(u^2+b^2)$ calculated with fluctuations in the frequency band $f\in[4.6,9.2]10^{-4}~\mathrm{Hz}$ (left panel); (b) the solar wind speed and the skewness of the radial velocity (central panels); and (c) the average magnetic field angle, $\theta_{BR}$, along with its standard deviation, $\Delta\theta_{BR}$ (right panels).
Most of the intervals have a large average cross-helicity (left panel), the solar wind speed is almost uniformly distributed with a slight dominance of slow streams (top central panel), and the average angle of the magnetic field is clustered around $50^o$ and $120^o$ (top right panel), the latter being more inclined with respect to the nominal direction of the Parker spiral (indicated by vertical dotted lines). 
The radial velocity fluctuations are mostly asymmetric, with the distribution of the skewness $SK(V_x$) having a maximum around $-0.4$ (central bottom panel), which indicates the presence of one-sided fluctuations as seen in the top panels of Figure~\ref{fig1a} (recall that $V_R=-V_x$). 
The standard deviation of the magnetic field angle $\Delta\theta_{BR}$ has a narrow distribution (bottom right panel), with a mean value of $20^o$, so that intervals contain basically no polarity inversion.

In each plot, the thick histograms refer to a subsample of intervals having $|\sigma_c|\ge0.7$, which represents $57$ per cent of the entire sample. 
Their average solar wind speed spans the entire range of the distribution, reflecting the presence of classical fast and Alfv\'enic streams, along with the slow and Alfv\'enic streams recently analysed in \citet{Damicis_al_2018}. 
The Alfv\'enic streams also contribute mostly to the asymmetry of the fluctuations in $V_x$.
The remainder of the population is made up of slow streams with small Alfv\'enicity and a flat distribution of the radial velocity skewness. 

In Figure~\ref{fig1c} we plot the distribution of the same quantities for the weak-expansion dataset.
The dataset contains mostly non-Alfv\'enic fluctuations (left panel), embedded in slow streams (top central panel), and the mean field is preferentially perpendicular to the radial direction (top right panel).
The distribution of $\Delta\theta_{BR}$ is broad and has an average value of $\sim30^o$ (left bottom panel). Qt variance with the strong dataset, the mean field direction varies substantially within a given interval.
Finally, the distribution of the skewness of the radial velocity has a maximum around zero and an important secondary peak at negative values. Asymmetric radial velocity fluctuations are also contained in this dataset (central bottom panel).
\begin{figure}    
               \includegraphics[width=0.450\textwidth,clip=,trim={0cm 1.51cm 0cm 0cm}]{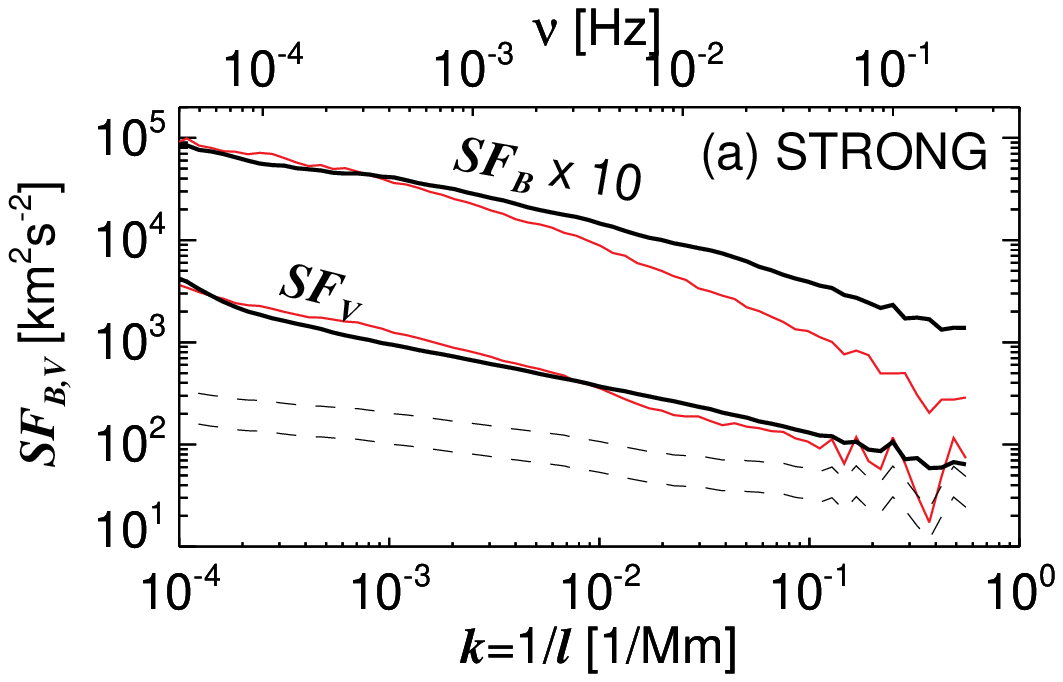}\\
               \includegraphics[width=0.450\textwidth,clip=,trim={0cm 0cm 0cm 1.cm}]{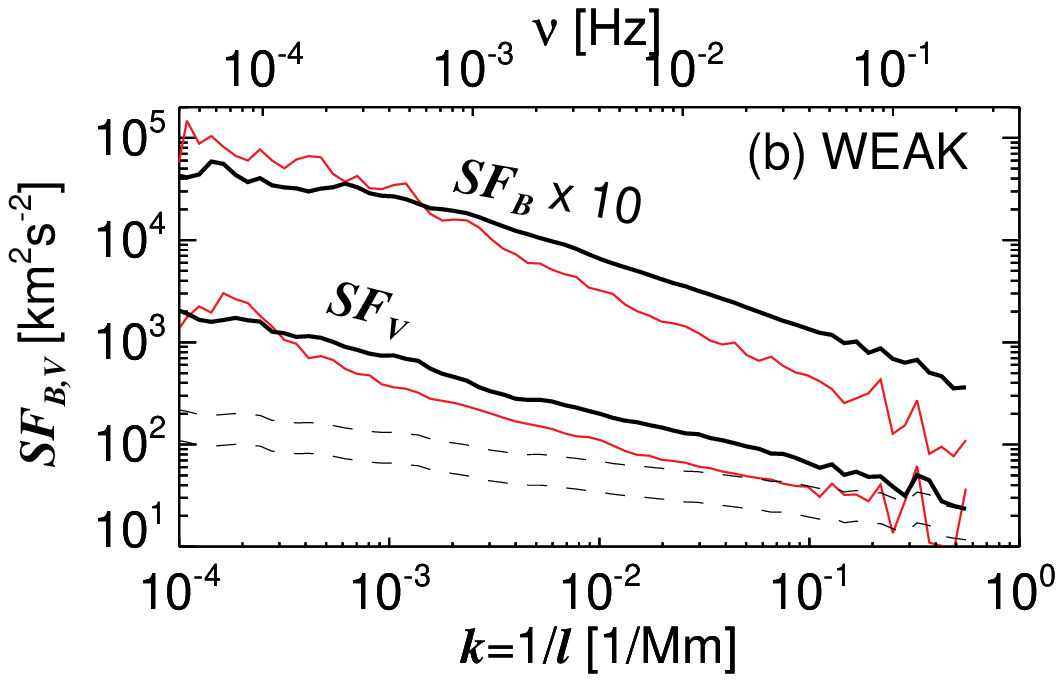}
              \caption{Raw axisymmetric $SF$ of velocity fluctuations ($SF_V$, bottom lines) and magnetic fluctuations ($SF_B$, upper lines) obtained by averaging, without any normalisation, the $SF_i$ of each interval in the strong (a) and weak (b) datasets (see definition in eq.\ref{axis-raw}). 
              The perpendicular and parallel $SF$ are drawn with thick black and thin red lines, respectively. The noise level on velocity fluctuations is plotted with a dashed line, along with twice its value (upper dashed line). In the top axis we indicate the frequencies corresponding to the length scale in the bottom axis, by assuming an average solar wind speed of $400~\mathrm{km~s^{-1}}$.}
   \label{fig7}
   \end{figure}

In the same figure, thick lines refer to distributions that are limited to a subsample with large cross-helicity, $|\sigma_c|\ge0.7$, which represents only the $27$ per cent of the entire sample.
This population of Alfv\'enic fluctuations contains fast and slow streams in about the same proportion, it has a slight asymmetry in the radial velocity fluctuations, but it conserves the properties related to the magnetic field direction: a perpendicular mean field with a large variation of its direction within an interval.

We conclude that the two datasets can be distinguished according to the mean magnetic field direction, its variability within an interval, and by the ratio $b_{\bot1}/b_{\bot2}$, rather than by the distribution of cross-helicity, although the strong-expansion dataset is mainly associated to Alfv\'enic fluctuations while the weak-expansion dataset contains mostly non-Alfv\'enic fluctuations.

\section{Results} 
\label{sec:results} 
We first analyse the anisotropy of magnetic and velocity structure functions, indicated in the next section with $SF_{B,V}$, respectively. We then study the 3D anisotropy of velocity fluctuations by relaxing the assumption of axisymmetry in the definition of the structure functions, $SF$ (we drop the subscript, the same analysis for magnetic fluctuations can be found in \citealt{Verdini_al_2018}).
In the following we will also measure the spectral index of $SF$. We recall that when the Fourier spectrum is a power-law in the whole range, $E\propto k^{-\gamma}$, than also the $SF$ is a power law, $SF\propto\ell^\alpha\propto k^{-\alpha}$, and its index is related to that of the Fourier spectrum by $\alpha=\gamma-1$. 

\subsection{Axisymmetric anisotropy} 
      \label{sec:axis} 

In Figure~\ref{fig7} we plot the axisymmetric raw structure functions (see eq.~\ref{axis-raw}) of the magnetic and velocity fluctuations for the strong and weak datasets, in the top and bottom panels, respectively, as a function of the wavenumber $k=1/\ell$. As a reference, in the top x-axis we also indicate the corresponding frequency scale, obtained by using an average solar wind speed of $400~\mathrm{km~s^{-1}}$. The magnetic structure functions, $SF_B$, are multiplied by a factor 10 to separate them from the velocity structure functions, $SF_V$, the black and red lines indicate the perpendicular and parallel $SF$, respectively.
For both datasets, $SF_B$ has a steeper slope in the parallel direction than in the perpendicular one, 
while for $SF_V$ the slope is approximately the same in the parallel and perpendicular directions. 
Without applying any normalisation before averaging among intervals belonging to the same dataset, $SF_V$ has a power-law index that is independent of $\theta_{BV}$. This is reminiscent of the results of \citet{Wang_al_2014},  but note that they averaged the slopes measured in seven fast streams intervals, so that a more appropriate comparison requires a proper normalization in our dataset (see below).

In the same figure we also plot, with a dashed line, an evaluation of the noise associated to the quantization of velocity measurement. Following \citet{Wicks_al_2013},
we first estimate the error on velocity measurements as the quadratic mean of the most probable value of the velocity increment at 3s in each interval (about $2~\mathrm{km~s^{-1}}$ for the radial component and $1.5~\mathrm{km~s^{-1}}$ for the other components), and then propagate this error in the definition of $SF$. 
The perpendicular $SF_V$ is always larger than the noise level, but only by a factor 2 at the smallest scales.
Instead, the noise becomes at least half of the signal in $SF_V$ at \textit{parallel} scales smaller than $8~\mathrm{Mm}$ for strong expansion and smaller than $25~\mathrm{Mm}$ for weak expansion, respectively. 
A scale of 10~Mm approximately corresponds to a frequency of $f=4~10^{-2}~\mathrm{Hz}$, which is close to the value estimated in \citet{Wicks_al_2013} for the noise to become important.

\begin{figure}    
               \includegraphics[width=0.450\textwidth,clip=,trim={0cm 1.5cm 0cm 0cm}]{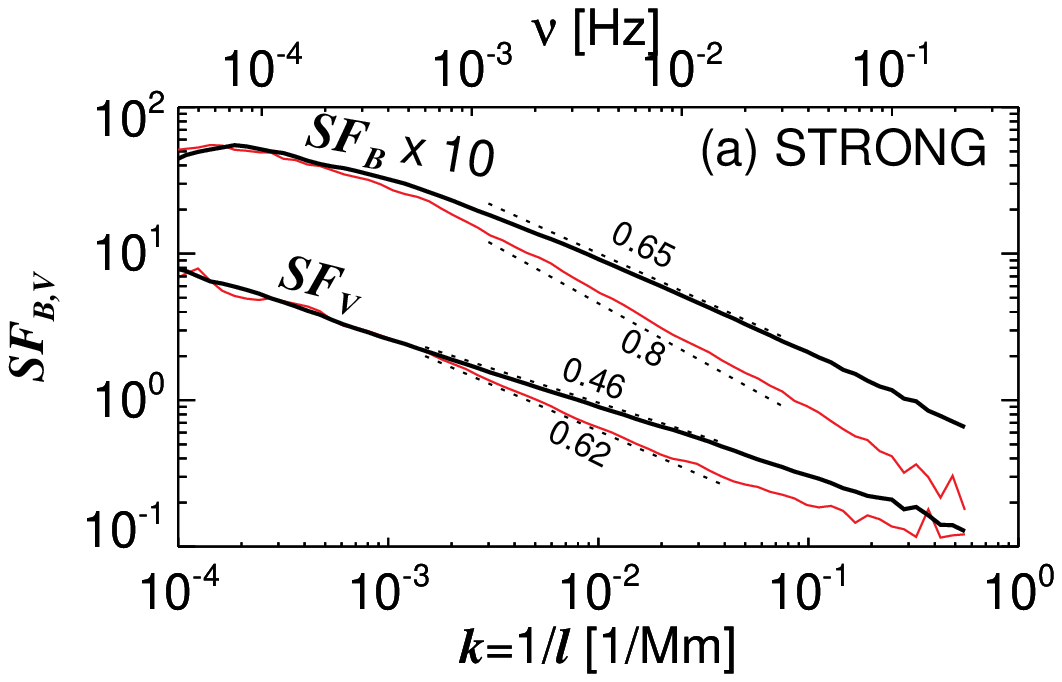}\\
               \includegraphics[width=0.450\textwidth,clip=,trim={0cm 0cm 0cm 1.cm}]{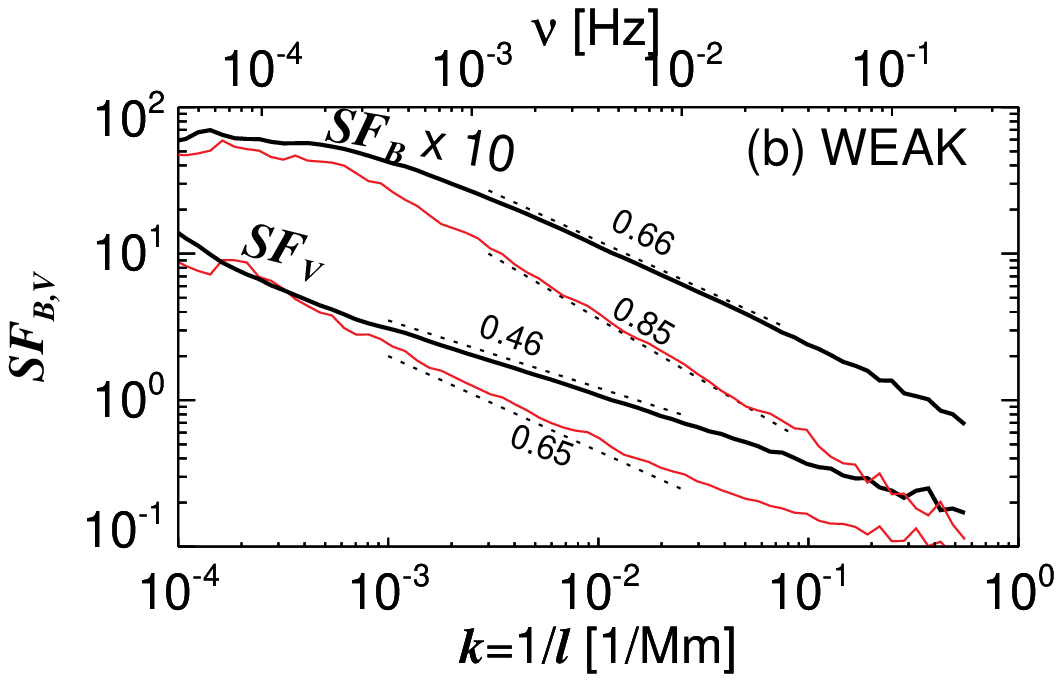}
              \caption{Normalized axisymmetric $SF$ of velocity fluctuations (see eq.~\ref{axis}) in the same format as in Figure~\ref{fig7}. Dotted lines are drawn as reference for the power-law scaling $SF\propto k^{-\alpha}$ with the index $\alpha$ indicated in the figure.
                      }
   \label{fig7b}
   \end{figure}
In Figure~\ref{fig7b} we plot again the axisymmetric $SF$ for magnetic and velocity fluctuations, this time applying a normalisation at scale $\ell^*\approx 100~\mathrm{Mm}$  (see eq.~\ref{axis}).
One obtains a clearer power-law behavior compared to the raw $SF$ in Figure~\ref{fig7}, and, more importantly, $SF_V$ now has a dependence on $\theta_{BV}$ in both datasets, with larger spectral indices in the parallel directions, in the range $k\in[2~10^{-3},4~10^{-2}]~\mathrm{Mm^{-1}}$ for strong expansion and $k\in[10^{-3},2~10^{-2}]~\mathrm{Mm^{-1}}$ for weak expansion. At larger $k$ the parallel $SF_V$ has approximately the same scaling of the perpendicular $SF_V$, possibly reflecting the small signal to noise ratio at those scales.
Note that a different scaling in parallel and perpendicular directions for both $SF_{B,V}$ shows up at larger scales in the bottom panel, indicating that the approximate criterion used to minimize expansion effects works fairly well. 

In the same figure we also indicate the values of the spectral index $\alpha$, measured by fitting $SF\propto k^{-\alpha}$ in the interval indicated by the dotted lines.
The spectral indices of the perpendicular $SF_{B,V}$ are the same independently of the strength of expansion, with a Kolmogorov-like index $\alpha_\bot\sim0.65$ for $SF_B$ and an Iroshinikov-Kraichan-like index, $\alpha_\bot\sim0.46$ for $SF_V$, very close to the average values found in observations for Fourier spectra.
The parallel spectral indices have a weak dependence on the strength of expansion, passing from $\alpha_\|\sim0.8$ to $\sim0.85$ for magnetic fluctuations, and from $\alpha_\|\sim0.62$ to $\sim0.65$ for velocity fluctuations when expansion is weaker. 
Note that at variance with \citet{Wang_al_2014} we obtain parallel $SF_V$ that are steeper than the perpendicular $SF_V$, but the parallel spectral index is much smaller than the value $1$ found by \citet{Wicks_al_2011}.

\subsection{Three-dimensional anisotropy} 
\label{sec:3Danis}
\begin{figure}    
               \includegraphics[width=0.980\linewidth,clip=,trim={0cm 1.5cm 0cm 0 cm}]{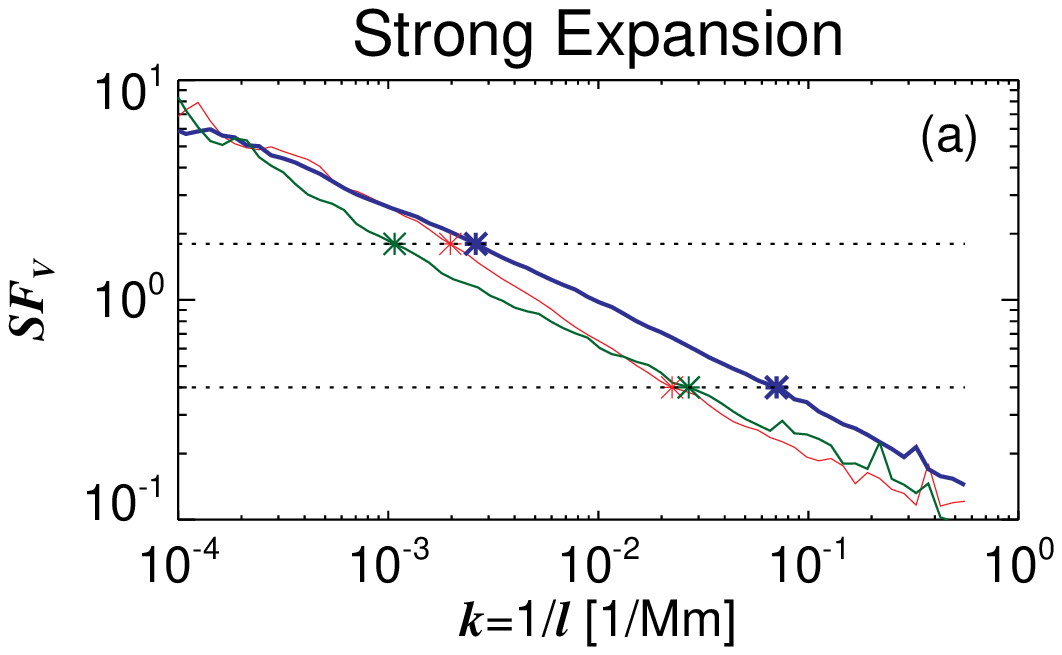}\\
               \includegraphics[width=0.980\linewidth,clip=,trim={0cm 0cm 0cm 0.72cm}]{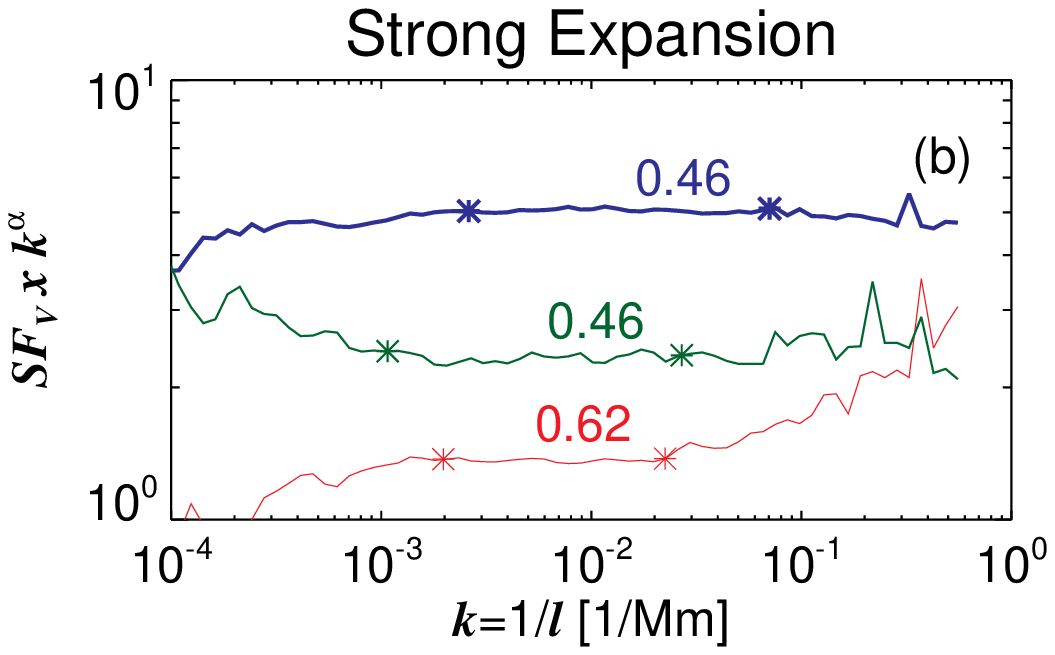}
              \caption{Strong expansion dataset. Panel (a): structure function of velocity fluctuations along the three orthogonal axes of the local reference frame (see Figure~\ref{fig:frame}): the perpendicular, displacement, and parallel directions are drawn with thick blue, green, and thin red lines respectively. Panel (b): the same three directions are now compensated by the power-law indices measured in the range $SF\in[1.5,5]$ (delimited by asterisks on each line) and indicated on top of each line. The lines are vertically shifted by an arbitrary factor for better visualisation
                      }
   \label{fig2}
   \end{figure}
In Figure~\ref{fig2}a we plot the structure functions of the velocity fluctuations, now indicated by $SF$, for the strong expansion dataset, in the three orthogonal directions, the perpendicular, displacement, and parallel directions ($\ell_\bot,~L_\bot$ and $\ell_\|$ in blue, green, and red, respectively), that are defined as:
\ba
SF(\ell_\bot)&\rightarrow&(85^o<\theta_B<90^o,85^o<\phi_{\delta B\bot}<90^o),
\label{per}\\
SF(L_\bot)&\rightarrow&(85^o<\theta_B<90^o,0^o<\phi_{\delta B\bot}<5^o),
\label{flu}\\
SF(\ell_\|)&\rightarrow&(0^o<\theta_B<5^o,0^o<\phi_{\delta B\bot}<90^o).
\label{par}
\ea
The parallel and perpendicular $SF$s have the same energy at large scales, a marker for strong expansion \citep{Verdini_Grappin_2015}. 
For $k\gtrsim10^{-3}~\mathrm{Mm}^{-1}$ the parallel $SF$ has a steeper slope and becomes subdominant at small scales, while the perpendicular and displacement $SF$s become parallel to each other.
At very small energies the parallel $SF$ (red line) starts to flatten, as already seen in the axisymmetric $SF$ in Figure~\ref{fig7b}a. 
The precise scaling laws in the three orthogonal directions are shown in Figure~\ref{fig2}b where the the $SF$ are compensated by the power-law index measured in the energy interval $0.4\lesssim SF\lesssim1.5$ (marked by asterisks in both panels and indicating our fiducial inertial range). 
The two perpendicular $SF$s have a slope of $0.46$ while the parallel $SF$ has a steeper slope, close to $0.62$.
The inequality $SF(\ell_\bot)>SF(L_\bot)$ suggests that power anisotropy is a consequence of the large-scale component anisotropy, $b_{\bot1}/b_{\bot2}<1$ seen in Figure~\ref{fig1}b. 
In fact, on the one hand, $SF(\ell_\bot)$ is measured when the local mean field direction is along $b_{\bot2}$, and $SF(L_\bot)$ is measured when the local field direction is along $b_{\bot1}$. On the other hand, because of the strong Alfv\'enicity, velocity fluctuations are expected to be almost incompressible and hence subject to the divergence-less constraint. Indeed also the their large-scale fluctuations have a component anisotropy with a ratio $v_{\bot1}/v_{\bot2}<1$ (not shown).

\begin{figure}    
               \includegraphics[width=0.950\linewidth,clip=,trim={0cm 1.5cm 0cm 0cm}]{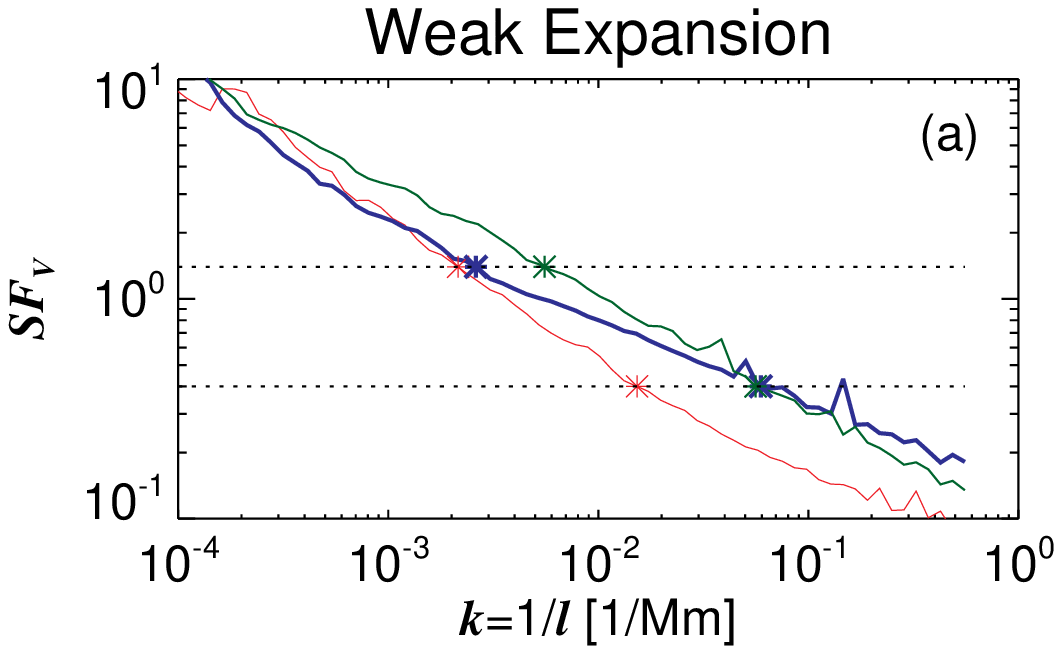}\\
               \includegraphics[width=0.950\linewidth,clip=,trim={0cm 0cm 0cm 0.72cm}]{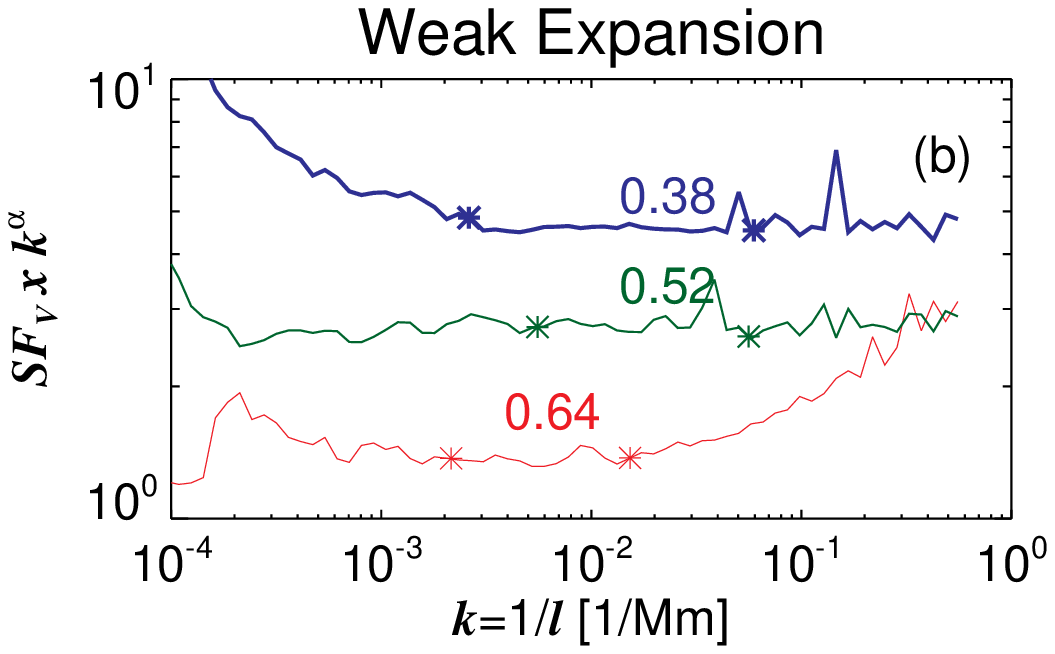}
              \caption{Weak expansion dataset. Same format as in Figure~\ref{fig2}.
                      }
   \label{fig4}
   \end{figure}

In Figure~\ref{fig4} we plot the $SF$ in the three directions for the weak-expansion dataset, also compensating them by the power-law index measured in the range $0.4\lesssim SF\lesssim 1.5$ (delimited by asterisks on the lines and representative of the inertial range). 
The displacement $SF$ is the most energetic at almost all scales, reflecting the large-scale component anisotropy $b_{\bot1}/b_{\bot2}>1$ already seen in Figure~\ref{fig1}b. This inequality is induced by the selection criterion for the weak-expansion dataset and implies that fluctuations are preferentially aligned with the radial direction, which is when $SF(L_\bot)$ is measured.
In the inertial range, the $SF$ has now three distinct slopes in the perpendicular, displacement, and parallel directions, with indices $0.38,~0.52$ and $0.64$, respectively. Below the bottom dashed line, i.e. for energies $\lesssim0.4$, the $SF$ is spiky, an indication that we are possibly reaching the noise in the measurements.

\begin{figure}    
               \includegraphics[width=0.220\textwidth,clip,trim={1cm 2cm 3cm 4cm}]{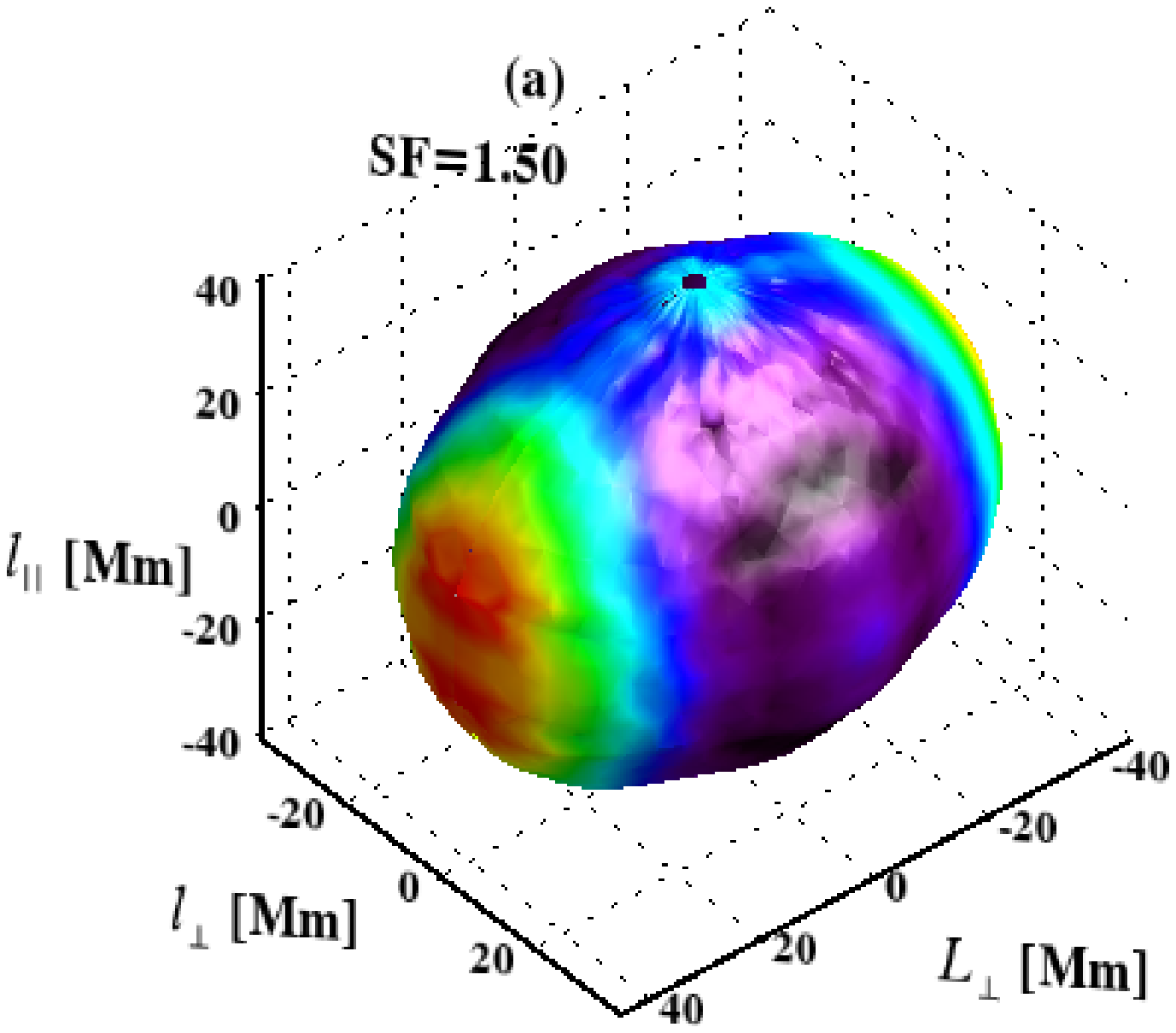}
               \includegraphics[width=0.220\textwidth,clip,trim={1cm 2cm 3cm 3.5cm}]{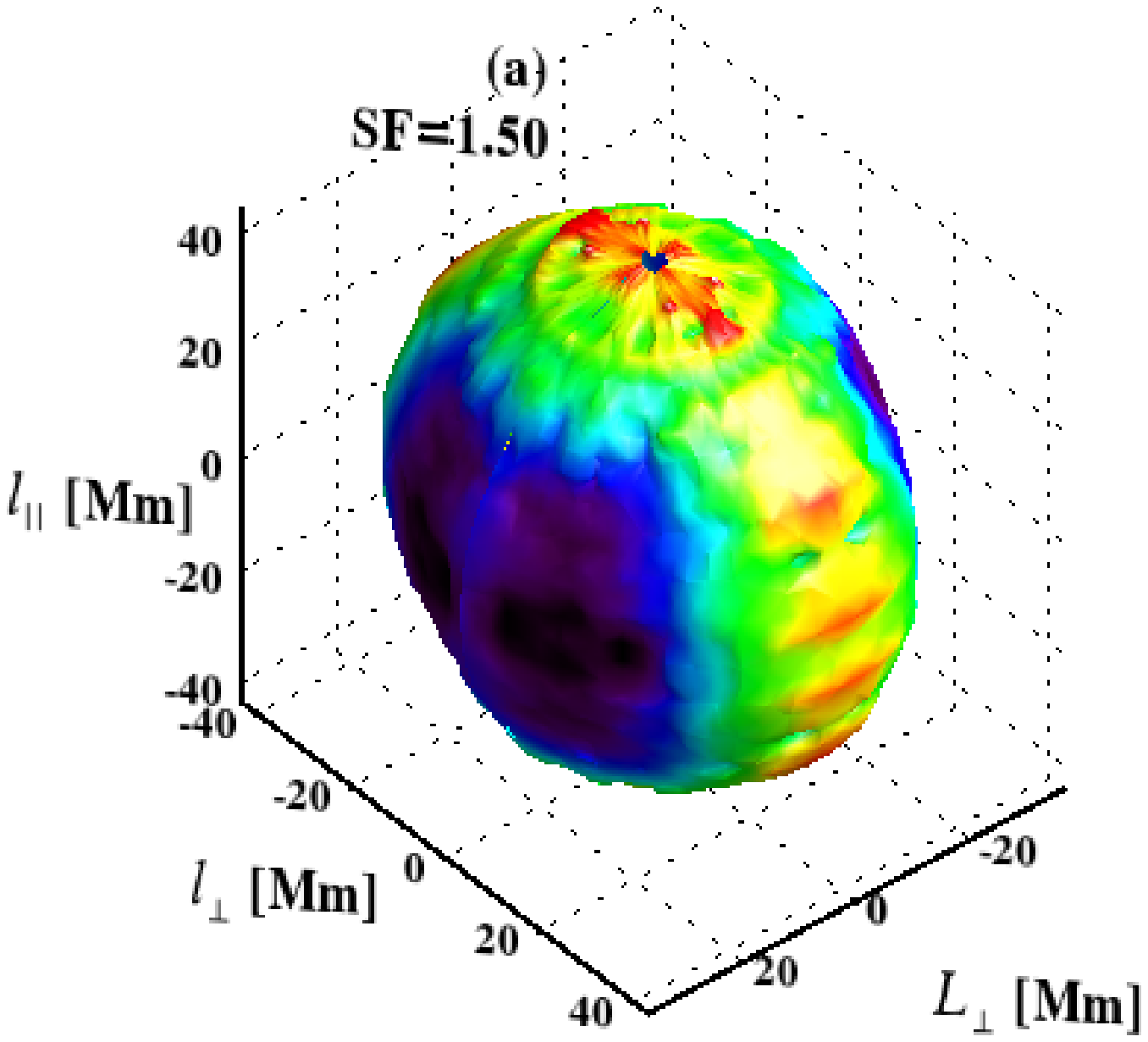}\\
               \includegraphics[width=0.220\textwidth,clip,trim={1cm 2cm 2.5cm 3cm}]{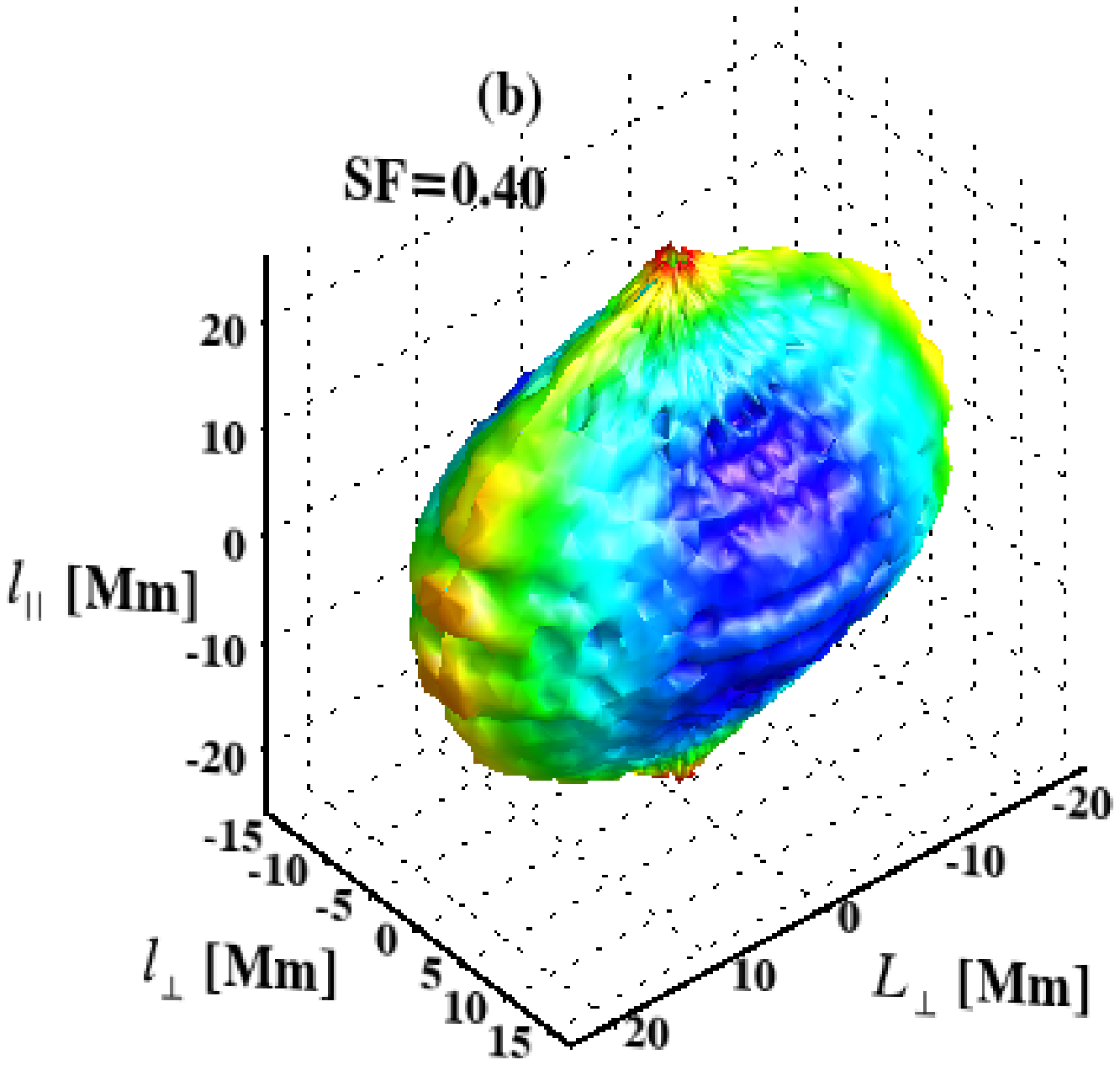}
                \includegraphics[width=0.220\textwidth,clip,trim={1cm 2cm 2.5cm 3cm}]{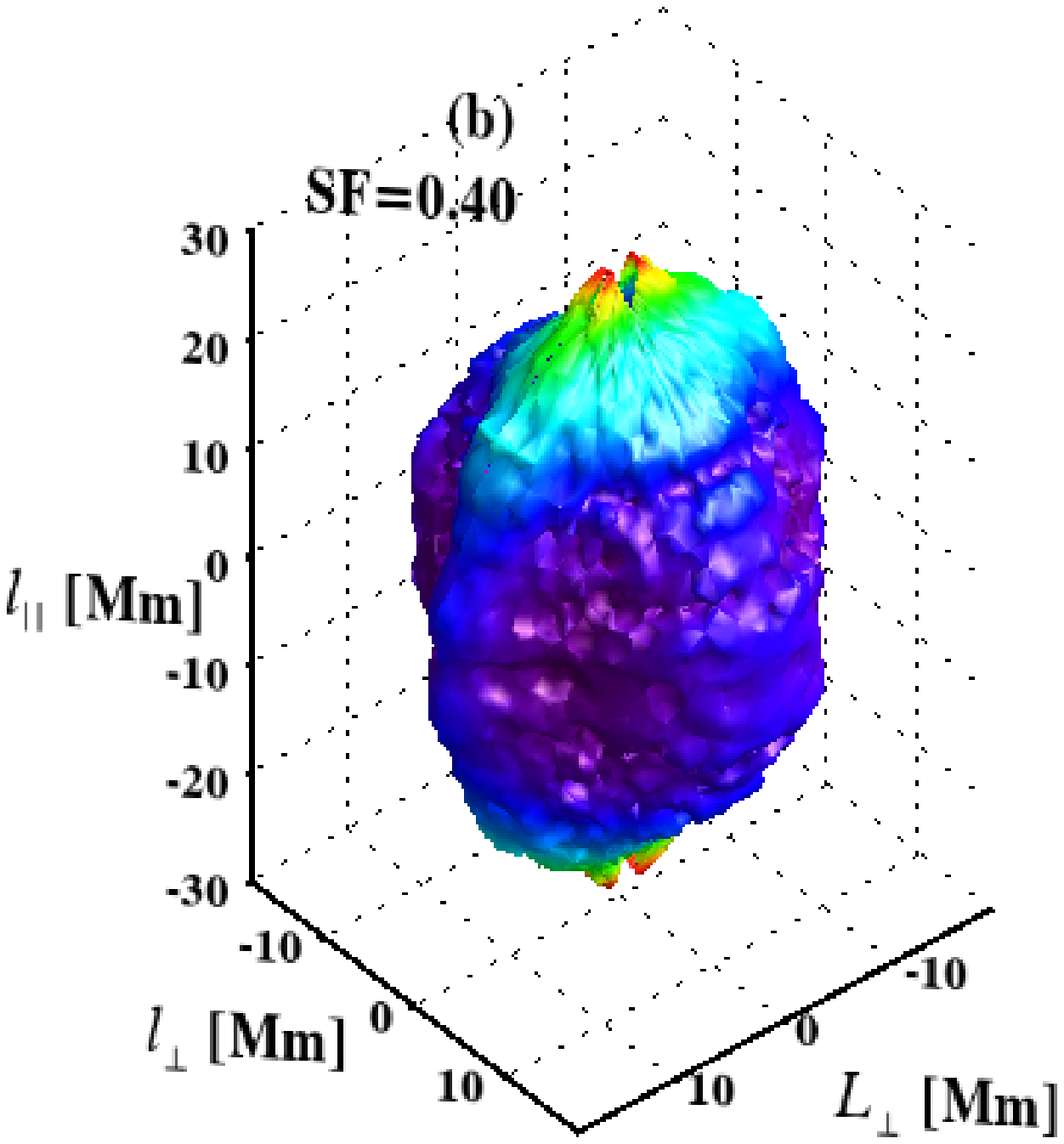}\\
               \includegraphics[width=0.220\textwidth,clip,trim={1.2cm 1.6cm 2.5cm 2.5cm}]{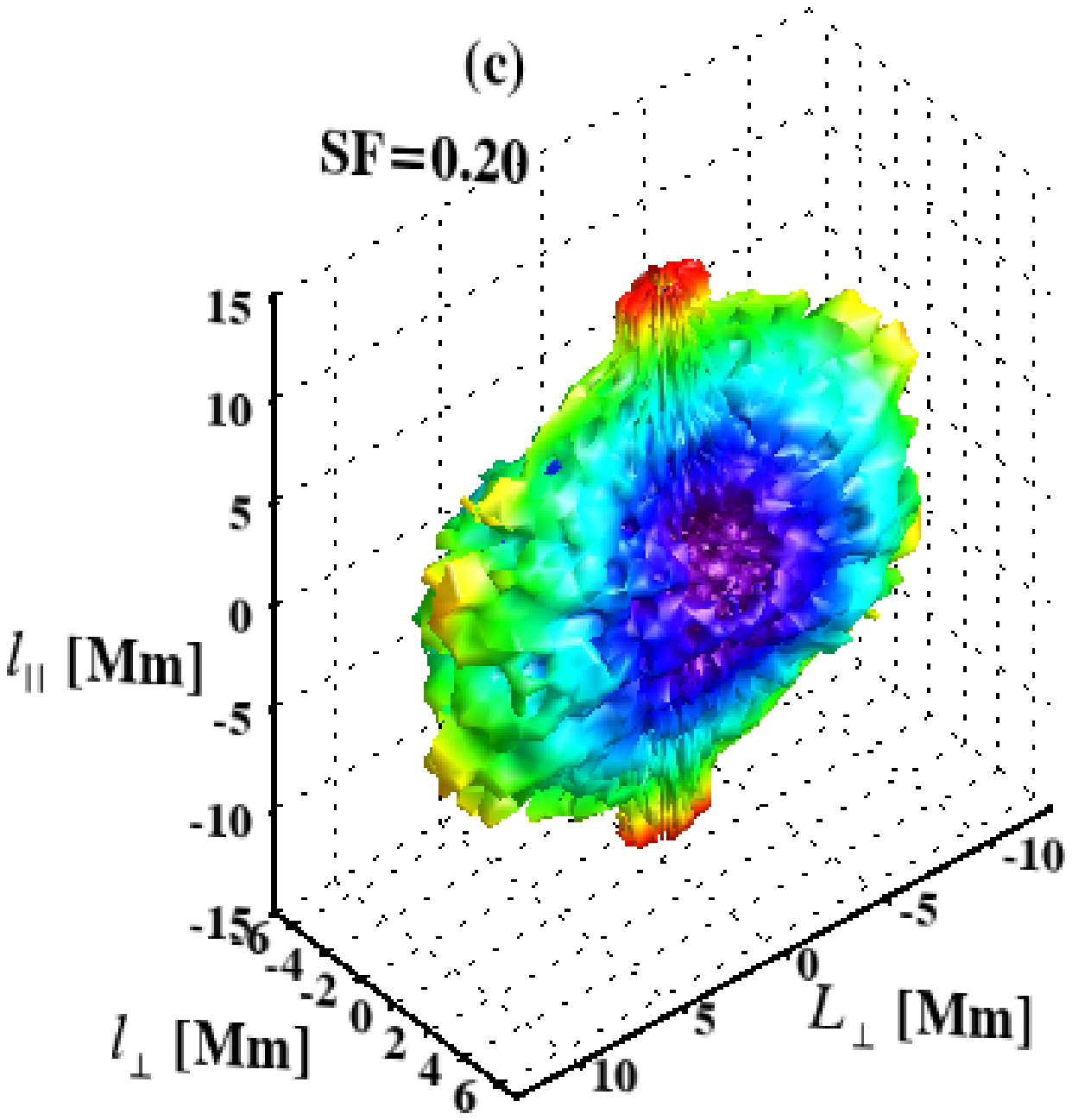}
                \includegraphics[width=0.220\textwidth,clip,trim={1cm 1cm 2.5cm 2cm}]{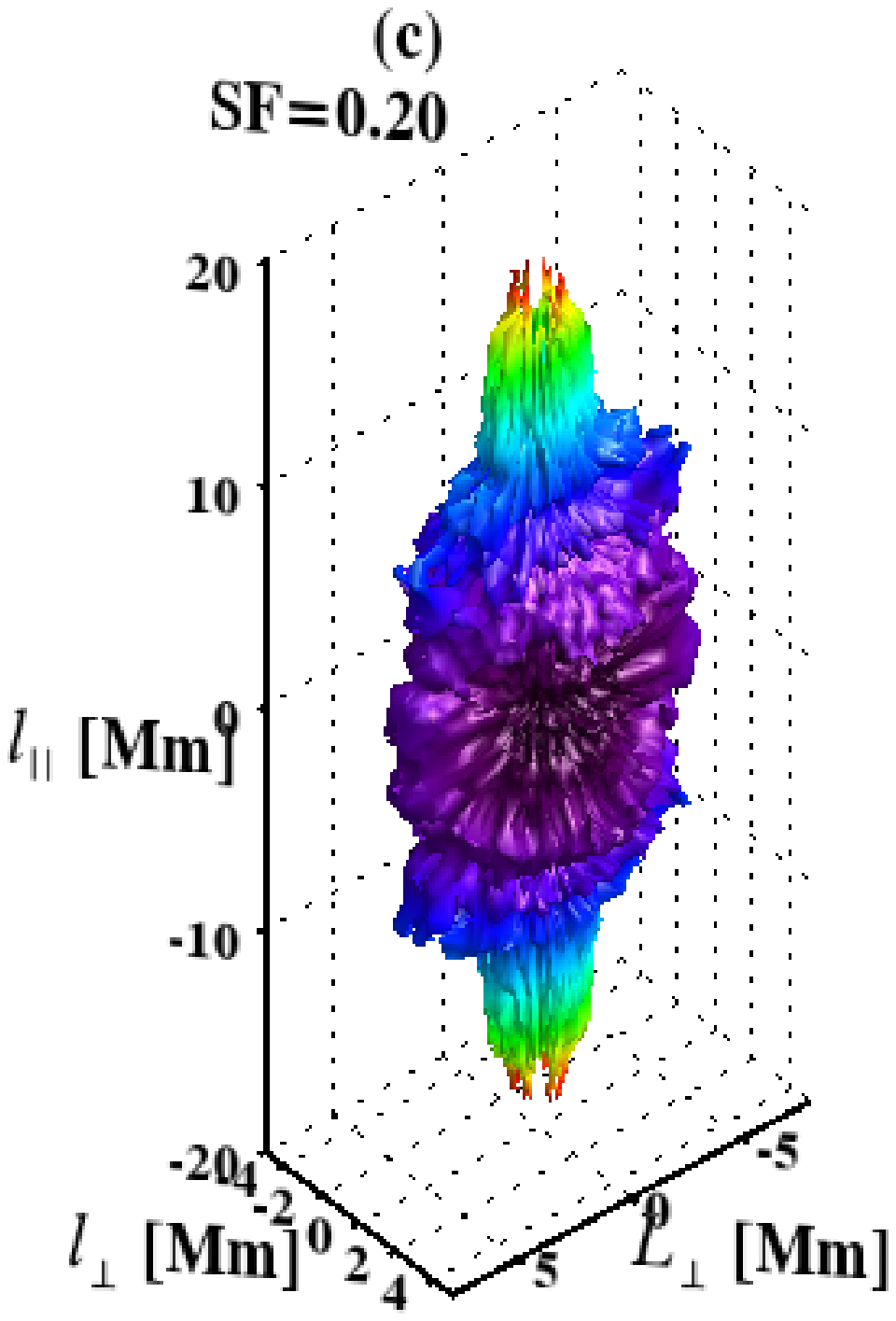}
              \caption{Strong expansion dataset (left) and weak expansion dataset (right). Isosurface of constant energy of the velocity structure function at three different levels, $SF\approx1.5,~0.4,~0.2$, in panels (a), (b), and (c) respectively. The colour is redundant, it indicates the distance from the centre for better visualisation.
                      }
   \label{fig3}
\end{figure}

The shape of turbulent eddies can be visualised in Figure~\ref{fig3}, where we draw the isosurfaces of constant energy of the $SF$ at three different levels, $SF\approx1.5,~0.4$, and $0.2$, corresponding to smaller and smaller scales.
The isosurface can be thought as the average shape of a turbulent eddy with given energy, as viewed in the varying frame attached to the local mean magnetic field.
For strong expansion (left panels), at the energy corresponding to the top boundary of the power-law interval in Figure~\ref{fig2}, the eddy is mostly elongated in the displacement direction and is about isotropic in the parallel and perpendicular directions (top left panel).
At intermediate energies (central left panel), corresponding to the bottom boundary of the power-law interval, the eddy is thinning in the perpendicular direction, which is now the smallest dimension, and becomes isotropic in the parallel and displacement directions.
At the smallest energy and scales (bottom left panel) the structure is again thinner in the perpendicular direction and maintains  an approximate isotropy in the parallel and displacement directions, although a larger elongation in the parallel direction begins to appear. 
 
In Figure~\ref{fig2}a one can see that the ratio $SF(\ell_\bot)/SF(L_\bot)\sim1.5$ at all scales. 
If $SF$ is measured in a fixed reference frame (global anisotropy) and the sampling direction is not aligned with the mean field direction, \citet{Saur_Bieber_1999} have shown that for an axisymmetric spectrum made of field-perpendicular fluctuations and wavevectors, the ratio of power in the the two components is $P_{\bot1}/P_{\bot2}=1/\alpha$, because of the divergence-less condition of magnetic fluctuations.
Although this argument applies only for measurements in a fixed reference frame, it strongly suggests that two local structure functions, $SF(\ell_\bot)$ and $SF(L_\bot)$, actually have the same power\footnote{We recall that $SF(\ell_\bot)$ is measured when the sampling direction is perpendicular to both the fluctuation and the mean field and it can be loosely associated to $P_{\bot2}$. Instead, $SF(L_\bot)$ is measured when fluctuations have non-vanishing projection in the sampling direction and can be loosely associated to $P_{\bot1}$. In addition, in this dataset, fluctuations are mainly Alfv\'enic so that also the velocity fluctuations are also expected to have vanishing divergence.}.  
If this is roughly true, all the eddy shapes shown in the left column should be squeezed in the $L_\bot$ direction. Thus, the small-scale shape that is similar to a disk would be consistent with a tube-like structure. 

In the right column of Figure~\ref{fig3}, we give the same 3D representation for the weak expansion dataset.
At the largest scales (top right panel), structures are again axisymmetric around the displacement direction, but the eddy is thinner along the axis of symmetry.  
At the bottom boundary of the inertial range (central right panel) one sees already a tendency to bi-dimensionalisation, with the main axis along the parallel direction and approximate axisymmetry.
At very small scales (bottom right panel), the eddy is clearly bi-dimensional with the main axis along the mean field and sheet-like, with a strong aspect ratio in the perpendicular plane, consistent with the different scaling in the perpendicular and displacement direction seen in Figure~\ref{fig4}. 
We conclude by noting that at the smallest scales (bottom panels), the eddies have dimensions comparable or smaller than 10~Mm, which is the scale at which the signal to noise ratio falls below 2 (see Figure~\ref{fig7}), thus casting doubts on their utility. We decided to show them because their shape is very similar to that one obtained with magnetic fluctuations (not shown, but see \citet{Verdini_al_2018} for the weak expansion dataset), which have much higher resolution and are not contaminated by noise.

\section{Summary and Discussion} 
      \label{sec:discussion} 
We have studied the local anisotropy of velocity fluctuations in two previously identified datasets in which expansion effects are expected to be small and large, respectively.
The selection criterion is $0<E<2$ for weak expansion and $2<E<10$ for strong expansion, with $E$ being the ratio of the energy in transverse and radial components of magnetic fluctuations calculated at 2h scale, $E=b^2_{tr}/b^2_{rad}$.
For strong expansion, the mean field direction is clustered around the Parker spiral and fluctuations are mainly Alfv\'enic, with a strong correlation between magnetic and velocity fluctuations.
For weak expansion, the mean field direction is preferentially perpendicular to the radial direction (a consequence of the selection criterion) and magnetic and velocity fluctuations are weakly correlated. 
When no distinction is made between the two field-perpendicular directions, that is, axisymmetry around the mean field is assumed, the two datasets have very similar distributions of the magnetic field compressibility, $b_\|/b_\bot<1$.
Also the axisymmetric local anisotropy as measured by structure functions, $SF$, is very similar. 

Let us denote with $\alpha_{\|,\bot}$ the power-law index $SF\sim k^{-\alpha}$ in the (local) parallel and perpendicular directions, respectively. 
For magnetic fluctuations, we find $\alpha_\bot\sim0.65,~0.66$ and $\alpha_\|\sim0.8,~0.85$ in the strong and weak-expansion datasets, respectively. 
Note that the difference in parallel and perpendicular spectral indices is slightly larger and appears at larger scales in the weak-expansion dataset, suggesting that the criterion used to limit expansion effects works fairly well despite being approximate (see \citealt{Vech_Chen_2016} for the relative contribution of expansion and the divergence-less constraint on the ratio $E$).
These results are consistent with previous findings \citep[e.g.][]{Horbury_al_2008,Podesta_2009, Luo_Wu_2010, Wicks_al_2010, Wicks_al_2011}, although the parallel index is significantly smaller than 1, a value expected for strong turbulence subject to the critical balance between the linear Alfv\'en time and the eddy turnover time \citep{GS95,B05,B06}. 
Small values of $\alpha_\|$ were obtained when intermittency was explicitly removed from the data \citep{Wang_al_2014} or when requiring the stationarity of the mean field direction \citep{Wang_al_2016}, which also indirectly limits intermittency. We have not measured intermittency in our datasets, and it seem unlikely that the selection criterion limits intermittency in the selected intervals.
Instead, our results are consistent with a small $\alpha_\|$ resulting from a non negligible contribution of the slab component \citep{VonPapen_Saur_2015}. Such component is mostly observed in fast wind streams with Alfv\'enic fluctuations, a population that is more abundant in the strong-expansion dataset, which actually displays a flatter parallel $SF$ than for weak expansion.

For velocity fluctuations the indices are generally smaller, with $\alpha_\bot\sim0.46,~0.46$ and $\alpha_\|\sim0.62,~0.65$ in the strong and weak-expansion datasets, respectively. 
At variance with \citet{Wicks_al_2013}, the parallel index is much smaller than 1 and it is consistent with the value found by \citet{Wang_al_2014} upon removal of intermittency from the data. As already said, it seems unlikely that intermittency is absent in our data and other explanations are needed. A possibility is that the parallel $SF$ is contaminated by noise in the velocity measurements that is different from the quantization noise estimated here, further work is needed to clarify this issue.

It is curious that $\alpha_\|$ and $\alpha_\bot$ of velocity $SF$ are related to each other by the critical balance between Alfv\'en time and a non-linear time, when the latter is based on velocity fluctuations. By equating the two timescales defined as $\tau_A=\ell_\|/V_A$ and $\tau_{NL}=\ell_\bot/v(\ell_\bot)$, respectively, and using the scaling $SF(\ell_\bot)\sim \ell_\bot^{1/2}$ seen in Figure~\ref{fig2}, one obtains $SF(\ell_\|)\sim\ell_\|^{2/3}$ which is very close to the measured parallel scaling.
This is suggestive of a turbulent regime in which the cascades of magnetic and kinetic energies proceed independently.
On the one hand, such a regime would contrast with the cascade of ideal invariants, since in incompressible MHD only the sum of the two energies is an invariant, and also with the strong Alfv\'enicity of fluctuations in the strong-expansion dataset. 
On the other hand, different spectral indices for the magnetic and velocity fluctuations are routinely found in numerical simulations, \citep{Milano_al_2001,Muller_Grappin_2004,Boldyrev_al_2012,Grappin_al_2016}, which indicates a sort of decoupling between the two fields, with a magnetic excess naturally developing at large scales and in the inertial range \citep{Grappin_al_1983,Muller_Grappin_2005,Boldyrev_Perez_2009}. 
Very recently it has been shown that when kinetic energy is injected at large scales, the transfer to magnetic energy stops to be effective at small scales in the inertial range, called the inductive range \citep{Bian_Aluie_2019}. 
In this range of scales, the cascades of kinetic and magnetic energy decouple and attain the same rate, thus supporting the above regime, although it is not clear yet if the independent cascades are associated to separate spectral indices for magnetic and kinetic energy.

When axisymmetry is relaxed, differences in the two datasets emerge. 
Consider first the properties of fluctuations at large scales, which are used to characterize the datasets.
Because of the small magnetic compressibility, $b_\|/b_\bot<1$, the selection criterion for weak expansion have two implications: (i) intervals have a mean field preferentially perpendicular to the radial direction; (ii) the field-perpendicular component lying in the $BR$ plane has a larger amplitude than the field-perpendicular component orthogonal to the $BR$ plane, $b_{\bot1} / b_{\bot2}>1$ (see Figure~\ref{fig1}b).
The selection criterion for the strong expansion dataset puts no constraint on the mean field direction and on ratio of the field-perpendicular components. The distribution of the magnetic field angle with the radial, $\theta_{BR}$, is thus clustered around the Parker spiral direction. However we find that $b_{\bot1} / b_{\bot2}<1$. This inequality is opposite to that of the weak-expansion dataset and can be understood as a consequence of the divergence-less constraint for $b$ when the mean field is not aligned to the (radial) sampling direction \citep{Saur_Bieber_1999}.
Thus, the two datasets have different large-scale variance anisotropy, $b_\|<b_{\bot1} < b_{\bot2}$ for strong expansion and $b_\|<b_{\bot2} < b_{\bot1}$ for weak expansion. 

The three-dimensional $SF$ of velocity fluctuations also have distinctive features in the two datasets. 
For strong expansion the perpendicular and displacement $SF$ have the same spectral index, $\alpha=0.65$, with smaller power in the displacement $SF$. This is due to the large-scale variance anisotropy $b_{\bot1}<b_{\bot2}$ and the Alfv\'enic character of fluctuations (see discussion in section~\ref{sec:3Danis}).
The constant ratio of the two perpendicular $SF$ at all scales suggests that the smaller power in the displacement $SF$ is an observational bias that hides an actual axisymmetry around the mean field as shown by \citet{Saur_Bieber_1999}.
Although the relation $SF(\ell_\bot)/SF(L_\bot)=1/\alpha$ holds only when anisotropy is computed in a fixed reference frame (global anisotropy), it is likely that a similar effect occurs for the local anisotropy analysed here. 

For weak expansion, we find a non-axisymmetric anisotropy: the perpendicular $SF$ is flatter than the displacement $SF$, their indices being $\alpha=0.38,~0.52$ respectively. However, because the large-scale variance anisotropy, $b_{\bot1}>b_{\bot2}$, the displacement $SF$ has more power than the perpendicular $SF$.
This implies that when axisymmetry is assumed, the perpendicular $SF$ takes the power-law index of the displacement $SF$, which explains why the axisymmetric anisotropy is the same in both datasets.
The resulting eddy shape, as measured by isosurfaces of constant energy, is also different in the two datasets. As a rule, at smaller and smaller scales, eddies become more elongated in the direction of the local magnetic field in both datasets. However, while for strong expansion the small-scale structure of eddies is consistent with an axisymmetric tube-like shape, for weak expansion it is more similar to a ribbon (although the latter correspondence is not strict because of the large power in the displacement $SF$)

The 3D anisotropy of magnetic fluctuations in the same datasets was analysed in \citet{Verdini_al_2018} who found similar properties. For strong expansion, the two perpendicular $SF$s have the same index $\alpha=0.65$, while for weak expansion $\alpha=0.53,~0.74$ in the perpendicular and displacement $SF$s, respectively.
The former are in agreement with earlier measurement in the fast polar wind \citep{Chen_al_2012}, and can be interpreted in light of the similar properties of the strong-expansion dataset: Alfv\'enic fluctuations with a large-scale component anisotropy, $b_{\bot1}<b_{\bot2}$. 
The latter are instead associated to the opposite component anisotropy, $b_{\bot1}>b_{\bot2}$, and to the mean magnetic field being perpendicular to the radial direction. 
In conclusion, we have shown that the measurement of the three-dimensional local anisotropy for both magnetic and velocity fluctuations is largely influenced by the underlying large-scale variance anisotropy of the magnetic fluctuations, as already pointed out in numerical simulations of MHD turbulence with expansion \citep{Verdini_Grappin_2015}.

Our results confirm that the difference in the magnetic and velocity spectral indices is a solid property that shows up whether we compare Fourier spectra, or parallel and perpendicular axisymmetric $SF$, or non-axisymmetric $SF$.
This is somehow paradoxical,  since magnetic and velocity fluctuations exhibit a strong coupling as measured by the cross helicity ($2\delta\vect{v}\cdot\delta \vect{b}/|\delta \vect{b}^2+\delta \vect{v}^2|$) or by the alignment of magnetic and velocity fluctuations ($\delta\vect{v}_\bot\cdot\delta \vect{b}_\bot/|\delta \vect{b}_\bot||\delta \vect{v}_\bot|$), but at the same time the two fields can have decoupled cascades \citep{Bian_Aluie_2019}.
A local measure of the alignment angle in the same datasets used here is reported in \citet{Verdini_al_2018e}. Although the progressive alignment with scales stops at relatively large scales, the angle remains small ($\sim23^o$) in the weak-expansion dataset \citep[see also][for similar results]{Podesta_al_2009, Wicks_al_2013}. Such an alignment suggests a strong (non-linear) coupling between velocity and magnetic fluctuations, which indeed have qualitatively similar anisotropies. However, this measure puts not constraint on the relative amplitudes of the two fields.
Instead, \citet{Chen_al_2013} found that for strongly Alfv\'enic intervals, the magnetic spectral index approaches that of velocity fluctuations with value $-3/2$. We find here a slope consistent with $-5/3$ in the strong-expansion dataset, where most of the intervals have $\sigma_c>0.7$, and the same index is reported by \citet{Chen_al_2012} that analysed fast streams with $\sigma_c\approx 0.6$. This indicates that only very large values of cross helicity ($|\sigma_c|>0.9$) cancel the difference in spectral index between velocity and magnetic spectra, that is, alignment alone is not enough and equipartition between the two fields must also hold.

Recent numerical simulations \citep{YangL_al_2017} and data analysis \citep{Wang_al_2014,Wang_al_2016,Bowen_al_2018} suggest that the differences in spectral indices are related to a stronger intermittency level in magnetic fluctuations, possibly due to the presence of pressure-balance structure in the solar wind. Intermittency correction should steepen the magnetic spectrum, but there is no particular reason for the resulting slope to be $-5/3$. Moreover, when intermittency is removed from data so as to obtain a monofractal behaviour of the exponent in higher order two-point correlations \citep{Salem_al_2009}, the spectral indices of magnetic and velocity fluctuations remains basically unchanged.
We have not analysed intermittency in our datasets, but in view of the above considerations on alignment and equipartition between velocity and magnetic fluctuations, our results seems to support an alternative scenario in which the spectral indices are regulated by a scale-by-scale equilibrium between the tendency to magnetic and kinetic equipartition (linear Alfv\'en effect) and the generation of magnetic excess (non-linear local dynamo) \citep{Grappin_al_1983,Muller_Grappin_2004,Muller_Grappin_2005,Grappin_al_2016}, with possibly a large-scale driver as expansion. However, again there is no particular reason for the indices of magnetic and kinetic spectra to be $-5/3$ and $-3/2$, respectively.
Understanding this property remains one of the most challenging achievement in solar wind turbulence.

\section*{Acknowledgements}
This work has been done within the LABEX PLAS@PAR project, and received
financial state aid managed by the Agence Nationale de la Recherche, as part of
the Programme ``Investissements d'Avenir'' under the reference
ANR-11-IDEX-0004-02.
RG and OA acknowledge support from Programme National PNST of CNRS/INSU co-funded by CNES.
\textit{Wind} data were obtained from CDAWeb (http://cdaweb.gsfc.nasa.gov).

This work was supported by the Programme National PNST of CNRS/INSU co-funded by CNES. 

\bibliographystyle{mnras}

\label{lastpage}
\end{document}